\documentclass[onecolumn]{IEEEtran}
\usepackage{times}
\usepackage{umoline}
\usepackage{times,amssymb,amsmath,amsfonts,float,nicefrac,color,bbm,mathrsfs,float,soul}
\usepackage{amsfonts}
\usepackage{latexsym}
\usepackage{amssymb}
\setstcolor{red}
\usepackage{comment}

\usepackage{theorem}
\usepackage{graphicx}
\usepackage{enumitem}

\usepackage{color,cite}
\usepackage[T1]{fontenc}
\usepackage{cmbright}

\newtheorem{theorem}{Theorem}
\newtheorem{definition}{Definition}
\newtheorem{proposition}{Proposition}

\newtheorem{lemma}{Lemma}

\newtheorem{example}{Example}
\newtheorem{remark}{Remark}

\newcounter{enumrom}
\renewcommand{\theenumrom}{(\roman{enumrom})}

\makeatletter
\renewcommand{\@endtheorem}{\endtrivlist}
\makeatother

\newcommand\remove[1]{}

\makeatletter
\renewcommand{\thefigure}{{\@arabic\c@figure}}
\renewcommand{\fnum@figure}{{\bf Figure\,\thefigure}}
\makeatother

\newcommand{\cA}{\mathcal{A}}

\newcommand{\cC}{\mathcal{C}}

\newcommand{\cH}{\mathcal{H}}

\newcommand{\cN}{\mathcal{N}}

\newcommand{\cS}{\mathfrak{S}}

\newcommand{\cX}{\mathcal{X}}

\newcommand{\bbs}{\boldsymbol{S}}

\newcommand{\mathset}[1]{\left\{#1\right\}}
\newcommand{\abs}[1]{\left|#1\right|}

\newcommand{\parenv}[1]{\left( #1 \right)}
\newcommand{\sparenv}[1]{\left[ #1 \right]}

\newcommand{\bal}[1]{\begin{align}\label{#1}}
\newcommand{\eal}{\end{align}}

\renewcommand{\le}{\leqslant}
\renewcommand{\leq}{\leqslant}
\renewcommand{\ge}{\geqslant}
\renewcommand{\geq}{\geqslant}

\renewcommand{\Bbb}{\mathbb}

\newcommand{\Cref}[1]{Co\-ro\-lla\-ry\,\ref{#1}}

\renewcommand{\Bbb}{\mathbb}

\newcommand{\N}{{\Bbb N}}

\newcommand{\R}{{\Bbb R}}

\newcommand{\E}{{\Bbb E}}

\newcommand{\ccap}{\mathsf{cap}}

\newcommand{\obeta}{\beta}
\newcommand{\ot}{\boldsymbol{t}}
\newcommand{\os}{s}
\newcommand{\obbs}{\bbs}
\newcommand{\oT}{T}
\newcommand{\wt}{h^{\ast}}
\newcommand{\bcomment}[1]{\textcolor{black}{#1}}

\outer\def\proclaim #1. #2\par{\medbreak
 \noindent{\bf#1.\enspace}{\sl#2\par}%
 \ifdim\lastskip<\medskipamount \removelastskip\penalty55\medskip\fi}

\begin{document}

\title{\textbf{Capacity of dynamical storage systems}}

\author{\IEEEauthorblockN{Ohad Elishco} \and\hspace*{.5in} \IEEEauthorblockN{Alexander Barg} 
}

\maketitle

{\renewcommand{\thefootnote}{}\footnotetext{

\vspace{-.2in}
 
\noindent\rule{1.5in}{.4pt}

This paper was presented in part at the 2019 IEEE International Symposium on Information Theory (ISIT), Paris, France, July 2019.

{O. Elishco is with Institute for Systems Research, University of Maryland, College Park, MD 20742, email ohadeli@umd.edu. His research is supported by NSF grant
CCF 1814487.

A. Barg is with Department of Electrical and Computer Engineering and Institute for Systems Research, University of Maryland, College Park, MD 20742 and also with Institute for Problems of Information Transmission (IITP), Russian Academy of Sciences, 127051 Moscow, Russia. Email: abarg@umd.edu. His research was supported by NSF grants CCF1618603 and CCF1814487.
}
}}

\maketitle

\begin{abstract}
We introduce a dynamical model of node repair in distributed storage systems wherein the storage nodes are subjected to failures according
to independent Poisson processes. The main parameter that we study is the time-average capacity of the network in the scenario where a 
fixed subset of the nodes support a higher repair bandwidth than the other nodes. The sequence of node failures generates random permutations of the nodes in the encoded block, and we model the state of the network as a Markov random walk on permutations of $n$ elements. As our main result we show that the capacity of the network can be increased compared to the static (worst-case) model of the storage system, while maintaining
the same (average) repair bandwidth, and we derive estimates of the increase.
We also quantify the capacity increase in the case that the repair center has information about the sequence of the recently failed storage nodes.
\end{abstract}
\section{Introduction}

\remove{In the last decade, distributed storage systems have been studied extensively. Their many applications 
motivated the authors of \cite{DimGodWUWaiRam2010} to formulate the problem as a coding problem and studying it from a coding theory perspective. The basic problem considers saving an encoded file over a network containing several storage units and retrieving it when needed. In practice, storage units (nodes) may fail. In order to maintain the ability to retrieve the file, the failed units are to be restored using the information stored in the network. This information (or a part of it) is downloaded by a data center and used to restore the failed node. This may cause extensive internal data transfer which affects the system. Hence, the main goal is to minimize the amount of internal data transfers while maintaining the ability to retrieve the file. As it turns out, there is a trade-off between the amount if data stored in each node and the bandwidth used for recovering a failed node. The two end points of this trade-off curve are of main interest in the literature and are called minimum storage regenerating (MSR) and minimum bandwidth regenerating (MBR) codes.}

The problem of node repair based on erasure coding for distributed storage aims at optimizing the tradeoff of network traffic and
storage overhead. In this form it was established by \cite{DimGodWUWaiRam2010} from the perspective of network coding.
This model was generalized in various ways such as concurrent failure of several nodes \cite{cadambe2013asymptotic}, 
heterogeneous architecture \cite{AkhKiaGha2010cost,SohChoYooMoo2018}, cooperative repair \cite{kermarrec2011repairing}, and others. 
The existing body of works focuses on the failure of a node (or several nodes) and the ensuing reconstruction process, but puts less emphasis on 
the time evolution of the entire network and the inherent stochastic nature of the node failures.
The static point of view of the system and of node repair leads to schemes based on the worst case scenario in the sense that 
the amount of data to be stored is known in advance, {the amount of data each node transmits is known,}  and the repair capacity is determined by the least advantageous state of the network. Switching to evolving networks makes it possible to define and study the average amount of data moved through the network to accomplish repair, {and may give a more comprehensive view of the system.}

Several models of storage systems have been considered in the literature. The basic model of \cite{DimGodWUWaiRam2010} assumes that the amount of data that 
each node transmits to the repair center is fixed. The analysis of the network traffic and storage overhead relies on \cite{AhlCaiLiYeu00} which quantifies the maximum total amount of data (or flow) that can arrive at a specific point, 
but does not specify the exact amount of data that each node should transmit at each time instant. 
To use the communication bandwidth more efficiently, we assume the amount of data that each node transmits changes over time, while the 
total amount of communicated information averaged over multiple repair cycles is fixed. 

A similar idea appears, although not explicitly, in \cite{SilGanWanAlvDah2014}, where the authors propose to perform
repair of several failed nodes within one repair cycle with the purpose of decreasing the network traffic. The decrease can occur
if the information sent over a particular link can be used for repair of more than one node, thereby decreasing the repair bandwidth.
This scheme, which the authors called ``lazy repair,'' views the link capacity as a resource in network optimization, which in general
terms is similar to the underlying premises of our study. A related, more general model of storage that accounts for time evolution of the system, given in
\cite{Luby17}, attempts to optimize tradeoffs between storage durability, overhead, repair bandwidth, and access latency. Coding for minimizing latency has been considered on its own in a separate line of works starting with \cite{Joshi12}. We refer to \cite{Badita19} for an overview of the literature where access latency is considered
in the framework of queueing theory.

To further motivate the dynamical model, recall that cloud storage systems such as Microsoft Azure or Google file system
encode information in blocks. The information to be stored is accumulated  until a block is full, and then the block is sealed: the information is encoded, and the encoded fragments are distributed to storage nodes \cite{khan2012rethinking,calder2011windows}. This implies that a storage node contains encoded fragments from several different blocks and that the sets of storage nodes corresponding to different blocks may intersect partially. Therefore, a storage node may participate in recovery of several failed nodes simultaneously, which implies that the capacity of the link between the node and the repair center can be considered as a shared resource.

In this work, we make first steps toward defining a dynamical model of the network with random failures. The prevalent system model assumes homogeneous storage under which the links from the nodes to the repair center all have the same capacity.
We immediately observe that the dynamical approach does not yield an advantage in the operation or analysis of this model. 
For this reason we study storage systems such that
the network is formed of two disjoint groups of nodes with unequal (average) communication costs, 
which was proposed in the static case in \cite{AkhKiaGha2010cost}. We show that, under the assumption of uniform failure probability of the 
nodes, it is possible to increase the size of the file stored in the system while maintaining the same network traffic. This means that, while in \cite{AkhKiaGha2010cost} 
the node transmits the same amount of data each time that there is a failure, in our model the same node will transmit the same amount of data in the average over a sequence 
of repair cycles (the time). In addition, we provide a simple scheme that increases the size of the stored file compared to the static model, and study state-aware dynamical 
networks in which the repair center has causal knowledge of the sequence of the failed nodes. The idea of time averaging is motivated by the assumption that the network exhibits some type of ergodic behavior whereby the expected capacity can be related to minimum cut averaged over time in a sample path of the network evolution. \bcomment{Since in our derivations we rely on the value of the minimum cut, in effect we are assuming ``functional repair'' of the failed nodes as opposed to the more stringent
requirement of exact repair \cite{DimGodWUWaiRam2010,Balaji2018}.}

In Section \ref{sec:prob}, we present the dynamical model and give a formal definition of the storage capacity. The evolution of the network is formalized as a random 
walk on the set of node permutations. Using this representation, we argue that it suffices to limit oneself to discrete time. We also prove
the basic relationship between the storage capacity of a continuous-time network and the time-average min-cut of the corresponding
discrete-time network (Sec.~\ref{sec:discrete}). \bcomment{This part relies on standard arguments related to the discrete-time Markov chain obtained by sampling a Poisson process at the times of change.} The main results of this paper are collected in Sec.~\ref{sec:static} where we derive estimates of the average capacity of the fixed-cost storage model. We examine two approaches toward estimating the capacity. The first of them is related to a specific transmission protocol while the second relies on an averaging argument. 
In Section \ref{sec:memory} we analyze state-aware networks and extend the ideas of the previous section to obtain a lower bound on their capacity.
Finally, in Section \ref{Sec:ext} we consider the case of different failure probabilities of the nodes and establish a partial result regarding
a lower bound on capacity.

\section{Model Definition}\label{sec:prob}

In this section we define a storage network that evolves in time and describe the basic assumptions that characterize this evolution. 
We also define a sequence of information flow graphs, which enables us to define the capacity of a randomly evolving network. 

\subsection{Evolution of the network}
A storage network is a set of data storage units that save information (``file'') with the purpose of being able to retrieve it at a later time. The file is partitioned into fragments placed on different storage drives or nodes of the system. Node failures occur regularly, and to
maintain the integrity of the data, the file is encoded using an erasure-correcting code.  This incurs a penalty in terms of both
the storage overhead and increased network communication and delay in the course of repair of the failed nodes. Once a node has failed,
the system initiates the reconstruction process in the course of which the centralized computing unit (CU) downloads information from 
a subset of functional storage nodes and performs the recovery of the data stored on the failed node. The amounts of data downloaded from 
 the different helper nodes to the CU vary over time, and are selected with the objective of minimizing the repair bandwidth.
 Thus, the sequence of node repairs is a time-dependent process which accounts for the time evolution of the network in terms
 of the information flow graph.
 
 Apart from node repair, the system also performs the operation of data collection (reading the file). This operation is performed by
a Data Collector ($DC$) which contacts storage nodes that allow the retrieval of the data. Since the file is encoded with an erasure-correcting code, the DC can retrieve the file by contacting a subset of the storage nodes. 

Let us give a formal description of our storage network model. A storage network is a pair $(\cN,\obeta)$ where $\cN$ is a triple $\cN=(V,DC,CU)$ in which $V$ is a set of $n$ nodes (storage units) $V=\mathset{v_1,\dots,v_{n}}$, $DC$ is the data collector node, and $CU$ is the centralized computing unit node. The real nonnegative vector $\obeta=(\beta_1,\dots,\beta_n)$ gives the maximum average amount of data 
communicated from $v_i$ for the node repair, and will be discussed in more detail below. 

Every node $v_i$, $i\in [n] \triangleq\mathset{1,2,\dots,n}$ has the ability to store up to $\alpha$ symbols over some finite alphabet $F$.  
To store a file of size $M$
\bcomment{we encode the file using} an $(n,k)$ code $\cC$. The coordinates of the codeword are vectors over $F,$ and each coordinate is stored
in its own storage node in $V.$ To read the file, the $DC$ accesses at least $k$ nodes, obtaining the information stored in them, and
retrieves the original file.

The time evolution of the storage network is related to a random process of node failures. We begin our study assuming that
time is continuous starting at $t=0,$ when the encoded file is stored in the network. 
The time instances $t_1,t_2,\dots$ indicate consecutive node failures. Let $\os=({s}_1,{s}_2,\dots)\in V^{\infty}$ be the sequence of failed 
nodes, where $s_j$ is the node that fails at time $t_j$. We assume that in order to restore the data to a failed node (reconstruct the node), 
the $CU$ contacts a group of storage nodes, called helper nodes, accesses some of the data stored on them, and uses this data to accomplish the
recovery. In this work we assume that $CU$ contacts all the nodes except the failed node, i.e., we assume that the number of helper nodes is
$n-1$. Further, we assume that the definition of the storage network includes a set of parameters $\beta_i, i=1,\dots,n,$
where $\beta_i$ is the maximum amount of information that is downloaded from $v_i$ to $CU$ for node repair, averaged over the time instances $t_i.$ 
Specifically, \bcomment{we define a sequence of functions $\mathset{h_j}_j,$ where $h_j:\cA_j\to \N,$ that determines the number of symbols that each node transmits for the recovery of the node $s_j.$ 
Thus, node $v_i$ provides $h_j(v_i)$ symbols of $F$ for the repair of node $s_j$.
It is assumed that $\limsup_{l\to \infty} 
\frac{1}{l}\sum_{j=1}^{l}h_j(v_i)\le \beta_i$ for all $i$.}
The case of $h_j(v_i)=\beta_i$ will play a special role, and
we introduce a notation for it: Let
    \begin{equation}\label{eq:h*}
        \wt_j(v_i)=\begin{cases} \beta_i &\text {for all }j: s_j\ne v_i\\
            0 &\text{for all }j: s_j=v_j.
             \end{cases}        
   \end{equation}   
We will also write $h,\wt$ to refer to the infinite sequences $\mathset{h_j}_j,
\{{\wt_j}\}_j$, respectively.

Note that by definition, the weight function $\wt$ does not achieve $\beta$ with equality. This is because $\wt_j(v_i)=0$ whence $s_j=v_i$. Thus, it is possible to increase the maximum file size that can be stored by taking $h_j(v_i)=({1+\frac{1}{n-1}})\wt_j(v_i)$. However, this increment in the file size will also increase the repair bandwidth. Although, for simplicity, the results in this paper are compared to the constraints $\beta$, it is straightforward to compare the results to the constraints $({1+\frac{1}{n-1}})\beta$.

Given $\cN$ and a sequence $\os=(s_1,s_2,\dots )$ of nodes, we define a sequence of directed graphs $\{\cX_j^{\os}\}_{j\in \N},$ called {\em \textbf{information flow graphs}}, where $\cX_j^{\os}$ corresponds to $t\in [t_j,t_{j+1})$ and is a subgraph of $\cX_{j+1}^{\os}$. When no confusion occurs, we will write $\cX_j$. The 
sequence of information flow graphs is a formalization of the notion
of a (time-evolving) information flow graph that appears in the foundational paper \cite{DimGodWUWaiRam2010}. For ease of description (and in accordance with \cite{DimGodWUWaiRam2010}), 
we introduce a new node $\tilde{v}$ which is called the source node.

\begin{figure}[h!]
  \centering
  \includegraphics[width=\linewidth]{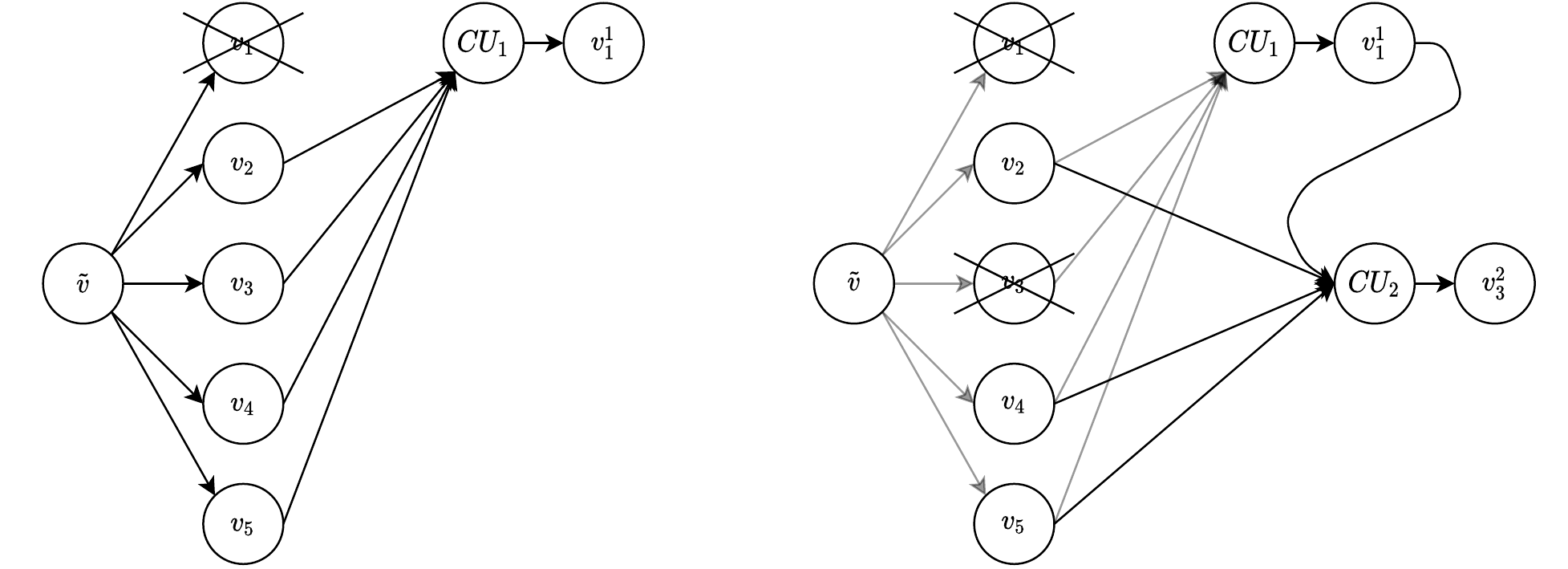}
  \caption{An illustration of $\cX_1^s$ (left figure) and $\cX_2^s$ (right figure) with $s=(v_1,v_3,\dots)$. 
  The set of active nodes for $j=1$ is $\cA_1=\mathset{v_1^1,v_2,v_3,v_4,v_5}$ and the set of active nodes for $j=2$ is $\cA_2=\mathset{v_1^1,v_2,v_3^2,v_4,v_5}$.}
  \label{fig:graph}
  \end{figure}

{\bf Definition: }
\begin{enumerate}[label= \arabic*:,leftmargin=.15in,labelindent=1em]
\item Let $V_0=V\cup \tilde{v}$ and put $\cX_0=(V_0,E_0)$, i.e., all the nodes in $\cN$ and the source node $\tilde{v}$, with edges  
\[E_0=\mathset{(\tilde{v},v_i) ~:~ i\in [n]}.\]
The nodes in the set $\cA_0:=V$ are called the {\em active nodes} of the graph $\cX_0$. We define $\cA_{-1}=\mathset{\tilde{v}}$.
\item Suppose that ${s}_1=v_{i_1}, i_1\in [n]$ and define a new node (newcomer) $v_{i_1}^1.$
The superscript $1$ implies that there was one failure and $v_{i_1}$ is the node that failed,  and the recovered node is $v_{i_1}^1$. 
The graph $\cX_1=(V_1,E_1)$ is formed as follows:
  \begin{gather*}
  V_1=V_0\cup \{CU_1, v_{i_1}^1\}\\ E_1=E_0\cup\{(v_j,CU_1),j\in[n]\backslash\{ i_1\}\}\cup (CU_1,v_{i_1}^1).
  \end{gather*}
  The set of active nodes of $\cX_1$ is defined as $\cA_1:=(\cA_0\backslash\{v_{i_1}\})\cup\{v_{i_1}^1\}$.
\item Suppose we are given the graph $\cX_{j-1}, j\ge 2.$ Suppose that ${s}_j=v_{i_j}$ and consider the corresponding node $v_{i_j}^{j'}$ in $\cX_{j-1}$ for some $j'<j$. 
\bcomment{The superscript $j'$ means that the node $v_{i_j}$ is the $(j')$th node that had been recovered (i.e., the superscript serves as a counter for the number of failures).}
Define a new node $v_{i_j}^j$ and define $\cX_j(V_j,E_j)$ as follows:
  \begin{gather*}
  V_j=V_{j-1}\cup\{CU_j, v_{i_j}^{j}\}\\
  E_j=E_{j-1}\cup\{(u,CU_j):u\in \cA_{j-1}\backslash\{v_{i_j}^{j'}\}\}\cup(CU_{j},v_{i_j}^j).
  \end{gather*}
  The set of active nodes of $\cX_j$ is defined as $\cA_j=(\cA_{j-1}\backslash\{v_{i_j}^{j'}\})\cup v_{i_j}^j.$ 
  We refer to Figure \ref{fig:graph} for an illustration.
  \end{enumerate}
The sequence of information flow graphs is an important tool used to represent the time evolution of the network. Each graph in 
the sequence accounts for a new node failure, and also records the information regarding all the past failures that occurred from the $t=0$ time. 
For a given $j$, the information for the repair of $s_j$ is communicated over the edges in the graph $\cX_j$, wherein the edge
$(v_i^{\ell},CU_j)$ carries $h_j(v_i)$ symbols of $F$, the index $\ell<j$ corresponds to the last instance when the node $v_i$ has failed.

We will sometimes write $(\cN,\obeta,\os,\ot,h)$ to denote a network $(\cN,\obeta)$ 
with the sequence of failed nodes $\os$, the sequence of failure times $\ot=(t_1,t_2,\dots)$, and a sequence of functions $\mathset{h_j}_j$.

In our model, the evolution of the network is random. \bcomment{Following earlier literature on storage networks, e.g., \cite{Luby17},} we
represent this evolution by assuming that the failure of each node is a Poisson
arrival process with rate $\lambda$, and these arrivals occur independently for different nodes. 
The interarrival time between two failures of a specific node $v\in V$ is an exponential random variable with pdf $\lambda e^{-\lambda t}$. Since 
node failures are independent, the overall rate of node failures in the system is a Poisson process with parameter $n\lambda$. This 
implies that we can formulate the network time evolution as follows. Let $(X_1,X_2,\dots)$ be a sequence of i.i.d. random variables with pdf 
$f_{X}(t)=n\lambda e^{-n\lambda t}$. Let $\oT=(T_1,T_2,\dots)$ be the sequence of failure times defined as $T_j=\sum_{i=1}^j X_i$ for $j\in\N
$ and let $\obbs=(\bbs_1,\bbs_2,\dots)$ be the sequence of failed nodes defined as a sequence of i.i.d. random variables 
distributed uniformly over $[n]$. Note that with probability zero the values $T_j$ can be infinite. Denote by $\mu_1$ the infinite 
direct power of the uniform distribution on $V$ and by $\mu_2$ the infinite power of the exponential distribution on $[0,\infty)$. 
We will assume that the sequence $(\obbs,\oT)$ is distributed according to $\mu_1\times\mu_2$.

\subsection{Data retrieval and network capacity}
To retrieve the file, the $DC$ contacts $k$ or more storage nodes and reads the information stored on them. Assume that the read request
occurs at time $t\in [t_j,t_{j+1})$ for some $j\in\N$. In the 
information flow graph $\cX_j$, the reading process amounts to introducing a new node $DC_j$ with at least $k$ incoming edges. Each edge originates in an active node. The 
set of these in-neighbors of $DC_j$ is denoted by $D_j\subseteq \cA_j$, $|D_j|\geq k$. 
The edges $\mathset{(v,DC_j): v\in D_j}$ have infinite weight. 
\bcomment{We denote by $k'$ the minimal number of active nodes from which the entire file can be retrieved.}

We are interested in the storage capacity of the network which is the maximum size of the file that can be stored in the network and be retrieved at any time while satisfying the average bandwidth constraints given by $\obeta$. Before defining the storage capacity, we need the following definition.
\begin{definition}
	Let $(\cN,\obeta,\os,\ot,h)$ be a storage network with a sequence of functions $\mathset{h_j}_j.$ The \textbf{$h$-capacity} of $\cN$, denoted by $\ccap_h(\cN)$,
	is the maximum file size that can be saved on the network $\cN$ and retrieved at any time. 
\end{definition} 

\begin{example}
	Let $(\cN,\obeta,\os,\ot,\wt)$ be a storage network and assume $\obeta_i=\beta_0$ for all $i\in [n]$. \bcomment{Note that in this case when a node fails all the active nodes transmit exactly $\beta_0$ symbols for the recovery process. It was shown in \cite{DimGodWUWaiRam2010} that the capacity in this case is equal to $\ccap_{\wt}(\cN)=\sum_{i=1}^{k'} \min\mathset{(n-i)\beta_0,\alpha}$. }
\end{example}
Note that the $\wt$-capacity expression contains a minimum. In order to simplify notation, we assume throughout that $\alpha$ is large enough, i.e., the storage nodes can contain any amount of information. This assumption allows us to remove the minimum in the capacity expression. In the previous example, we obtain that $\ccap_{\wt}(\cN)=\sum_{i=1}^{k'}(n-i)\beta_0$.

We now define the storage capacity. 
\begin{definition}
	Let $(\cN,\obeta,\os,\ot)$ be a storage network and let $\cH$ denote the set of all sequences of functions $h$ that satisfy the constraints given by $\obeta$. The \textbf{storage capacity} (or just capacity) of $(\cN,\obeta,\os,\ot)$, denoted by $\ccap(\cN)$, is defined as 
	\[\ccap(\cN)=\sup_{h\in \cH} \ccap_h(\cN).\]
\end{definition} 

The random evolution of the network makes the sequence of failed nodes a sequence of random variables which we henceforth denote by $\obbs$. This makes $\ccap(\cN)$ a random variable as well. As such, we will analyze the expected value of the storage capacity which is defined as follows.
\begin{definition}
	Let $(\cN,\obeta,\obbs,\ot)$ be a (random) storage network. The \textbf{expected capacity} is defined as 
	\[\overline{\ccap}(\cN)\triangleq \E\sparenv{\ccap(\cN)}.\] 
\end{definition}

For any realization $\os$ of $\obbs$, the storage capacity of a network can be calculated using the sequence of information flow graphs $\mathset{\cX_j}_j$.  Indeed, let $(\cN,\obeta,\os,\ot,h)$ be a storage network with a corresponding sequence of information flow graphs $\mathset{\cX_j}_j.$ 
For a time $t\in [t_j,t_{j+1}]$, $j\in \N$, let $D_t$ denote a selection of $k'$ nodes from $\cA_j$ (from which the entire file can be retrieved). 
As shown in \cite{DimGodWUWaiRam2010}, the storage capacity of the network is equal to the minimum weight of a cut between $\tilde{v}$ and $D_t$. 
In other words, the maximum file size that can be reliably stored in the network and retrieved at time $t$ is equal to the minimum weight of a cut between $\tilde{v}$ and $D_t$. 
\bcomment{A cut between $\tilde{v}$ and $D_t$ is a partition of the vertices into two sets where $\tilde{v}$ is contained in the first set and the $k'$ nodes in $D_t$ are contained in the second set. 
The weight of the cut is the sum of the weights of the edges from the first set to the second set. 
Therefore, if we can find a weight function $h$ such that for every time instance $t$ and for every selection $D_t$ of $k'$ active nodes a minimum cut between $\tilde{v}$ to $D_t$ is bounded below by a constant $C$, then $C$ is the maximum file size that can be saved in the network. Our goal is to bound $C$ by specifying a weight function $h$ that obeys the restrictions given by the average bandwidth of the nodes.}

In this work, we consider the time-average minimum cut (as defined below) and hence we define the minimum cut accordingly. 
\bcomment{\begin{definition}
    Let $C_t^h (D_t)$ denote the value of the minimum cut in $\cX_j$ between 
$\parenv{\bigcup_{i=-1}^{j-1} \cA_i}\setminus \cA_{j}$ and $D_t$ under the weight assignment $h$. 
Further, let $C^h_t$ denote the minimum cut over all selections $D_t$, 
   \begin{equation}\label{eq:mincut}
C^h_t=C^h_t(\cA_j)\triangleq \min_{D_t\subseteq \cA_j,\; |D_t|= k'} \mathset{C^h_t(D_t)}.
   \end{equation}
   When $h=\wt,$ we will sometimes write $C_t$ instead of $C^{\wt}_t$.
\end{definition}}
In this definition we again assume that the $DC$ is not aware of the state of the network, i.e., the order of the failed nodes, and the minimum accounts for the worst case. If $DC$ can choose which nodes to contact, the minimum should be replaced with a maximum.

\begin{definition}
	Let $(\cN,\obeta,\os,\ot,h)$ be a storage system. Define the \textbf{average cut} as
	\[C_{\mathrm{avg}}^h(\cN)\triangleq \limsup_{t\to\infty} \frac{1}{t}\int_0^t C^{h}_{\tau}\;\mathrm{d}\tau.\]
\end{definition}
Note that the average cut is a function of $\os$. Hence, if the network is random $\os=\obbs$, then the average cut is a random variable.
As shown in the following lemma, for a storage system $(\cN,\obeta,\os,\ot,\wt)$, the average cut $C_{\mathrm{avg}}^{\wt}(\cN)$ can be used to bound below the capacity of 
the network $\cN.$
\begin{lemma}
	\label{lem:lower_bound_iccap}
Let $(\cN,\obeta,\os,\ot)$ be a storage system. Assume that for any $n-k'$ nodes $v_{i_1},\dots,v_{i_{n-k'}}\in V$ and
all failure times $t=t_1,t_2,\dots$
	   \begin{equation}\label{eq:sum}
	   \sum_{j=1}^{n-k'}\beta_{i_j}\geq \max_{D_t\subseteq \cA_t, |D_t|=k'} \abs{C_t^{\wt}(D_t)-C_{\rm avg}^{\wt} }.
	   \end{equation}
	If every node fails infinitely often, then 
	   \begin{equation}\label{eq:ge}
	\ccap(\cN)\ge C_{\mathrm{avg}}^{\wt}(\cN).
	\end{equation}
\end{lemma}
{\em Remark:} Eq.~\eqref{eq:ge} in effect states that there exists a weight assignment $h\in \cH$ such that
$\ccap_h(\cN)$ is at least the size of the average cut under $\wt.$ 
\vspace*{.1in}\begin{IEEEproof}
	We prove the lemma by describing an algorithm for weight selection. In order to simplify the notation, for $t\in [t_j,t_{j+1})$ we use the subscript $j$ instead of $t$, 
for example, we write $C_j$ instead of $C_t$. Let $j_1$ be the first occurrence when $C_{j_1}<C_{\rm{avg}}^{\wt}.$ If $j_1$ is infinite, then there is nothing 
to prove; 	otherwise, for $j<j_1$ take $h_j=\wt_j$. 
Let $v_i$ be the node that failed at time $t\in [t_{j_1},t_{j_1+1}).$
For every \bcomment{$v_{\ell}\in \cA_j$ ($v_{\ell}\neq v_i$)} define $h_{j_1}(v_{\ell})=(1+\varepsilon_0)\wt_{j_1}(v_{\ell})$ for $\varepsilon_0>0$ such that the minimum cut $C_{j_1}^{h}(\cN)=C_{\rm avg}^{\wt}(\cN)$, \bcomment{i.e., every active node transmits more information to enlarge the minimum cut to $C_{\rm avg}^{\wt}(\cN)$.}
For time $t_{j_1+1}$ again find $\varepsilon_1$ such that for every $v_{\ell}\in\cA_{j_1+1},$ taking $h_{j_1+1}(v_{\ell})=(1+\varepsilon_1)\wt_{j_1+1}(v_{\ell})$ yields $C_{j_1+1}^{h}(\cN)=C_{\rm avg}^{\wt}(\cN)$. 
Note that $\varepsilon_1$ can be positive or negative. 
Continue this way to define $h_{j_1+r},$ and the respective values of $\varepsilon_r, r\geq 2.$ 

	By construction, the average number of symbols that node $v_i$ transmits is $\beta_i$. Moreover, at any time instance $t_{\ell}$, the file can be reconstructed from any selection $D_{\ell}$ of $k'$ storage nodes. Hence, the minimum cut between $\tilde{v}$ and $D_{\ell}$ 
	(the storage capacity) is at least $C_{\mathrm{avg}}^{\wt}(\cN)$.  
\end{IEEEproof}

\vspace*{.1in}  \bcomment{To explain assumption \eqref{eq:sum} note the following. Suppose a node was one of the 
nodes selected by $DC_{t-1},$ and suppose that it fails at time $t-1.$ 
If the recovered node is selected by $DC_{t}$, the weight of the cut between the source and the selected nodes is affected only by the $n-k'$ in-edges that connect the nodes not selected by $DC_{t}$ to the failed node.
Assumption \eqref{eq:sum} implies that for every selection $D_t$ of the $k'$ nodes and for every $r\geq 1$, it is possible to find an $\varepsilon_r\ge -1,$ which ensures the consistency of the transmission. }
The physical interpretation of the assumption given in \eqref{eq:sum} is that any set of $n-k'$ nodes contain more ``new  information'' than the information unavailable due to the node failure.  Throughout the paper we assume that \eqref{eq:sum} holds true.

We will establish a more detailed lower bound on $\ccap(\cN)$ in terms of $C_{\rm avg}^{\wt}(\cN)$ in Sec.~\ref{sec:discrete} below.


\subsection{Network evolution as a sequence of permutations}
In this subsection we define a set of permutations related to the sequence of failed nodes $\os=(s_1,s_2,\dots).$ 
Each node in $\cA_j$ is denoted by $v_{i_{j'}}^{j'}$ for some $j'\leq j$, and it can be identified with the node $v_i\in V$. 
Therefore, if at some point $t_{j_0}$ all the nodes have failed at least once, then for $j\geq j_0$ the order in which the nodes in $\cA_j$ 
have failed can be identified with a permutation of the set $[n]$. 
For example, if $\cA_j=\mathset{v_1^{j_1},v_2^{j_2},\dots,v_n^{j_n}},$ 
then the corresponding permutation $\pi$ is such that $\pi(i)\leq \pi(\ell)$ iff $v_i^{j_i},v_{\ell}^{j_{\ell}}\in\cA_j$ with $j_i\leq 
j_{\ell}$. Below we identify $V$ and $[n]$ and consider $\pi_t, t\ge t_{j_0}$ as a permutation of either of these sets as appropriate. We 
denote by $\cS_n$ the set of all permutations of $[n]$. The permutations $\pi_t$ are associated with the sequence of the information flow
graphs $(\cX_j)$, and we call $\pi_t$ the associated permutation (at time $t$).  Note that the associated permutation $\pi_t\in\cS_n$ corresponds to the order of the $n$ most recent node failures. Hence, for $t\in[t_j,t_{j+1})$ we will sometimes write $\pi_j$ instead
of $\pi_t$ to refer to the associated permutation at time $t$. Observe that the minimum cut $C_t(\cA_j)$ is a function of the associated
permutation $\pi_t$ for every $t\in [t_j,t_{j+1})$ and $j\geq j_0,$ so we can write $C^h_t(\cA_j)=C^h_t(\pi_t)$.

It is possible to obtain $\pi_t, t\geq t_{j_0}$ from $\os$ by considering only the last appearance of each node as seen in the next example.
\begin{example}\label{example1}
	Assume that $|V|=5$ and assume that $\os=(v_1,v_2,v_3,v_4,v_5,v_2,v_1,v_5,\dots)$ with $\ot=(1,2,3,\dots)$. Then $\pi_t, t\in [1,5)$ is not defined and $\pi_t=(v_1,v_2,v_3,v_4,v_5)=id$ for $t\in[5,6)$ since all the nodes had failed by $t=5$. At $t=6$ the node $v_2$ fails again, hence the new order is given by $\pi_t=(v_1,v_3,v_4,v_5,v_2)$ for $t\in[6,7)$, i.e., $v_{\pi_t(1)}=v_1, v_{\pi_t(2)}=v_3$ and so on. This is because the second node appears twice in $\os_1^6$ and we consider only the last appearance. Following the same reasoning, $\pi_t=(v_3,v_4,v_5,v_2,v_1)$ for $t\in[7,8)$ and $\pi_t=(v_3,v_4,v_2,v_1,v_5)$ for $t\in[8,9)$. 
\end{example}
We remark that if $\pi_t$ is an associated permutation, then $v_{\pi_t(i)}$ denotes the node that appears in the $i$th location and $\pi_t^{-1}(i)$ denotes the location 
of the node $v_i.$ 
In Example \ref{example1}, we have $\pi_6(2)=3$ since $v_3$ occupies the second position, and $\pi_6^{-1}(2)=5.$ 
\bcomment{Although $\pi_t$ is a function from $[n]$ to $[n]$, for a node $v_i$ we will often write $\pi^{-1}(v_i)$ to denote $\pi^{-1}(i)$. 
In Example \ref{example1}, we have $\pi_6^{-1}(v_2)=5$.}

Now suppose that the evolution of the network is random and let  $t_{j_0}$ be the time by which all the nodes fail at least once. The next lemma shows that such $j_0$ exists 
almost surely and that the associated permutation $\pi_{t_{j_0}}$ is uniformly distributed on $\cS_n.$

\begin{lemma}
	\label{lem:finitetime}
	Let $(\obbs,\oT)=\parenv{(\bbs_i,T_i),\; i\geq 1}$ be an infinite sequence distributed according to $\mu_1\times\mu_2$. Then almost surely, there exists a finite $t_0\in\R$ such that all the nodes have failed at least once by $t_0$. Moreover, $\pi_{t_0}$ is distributed uniformly on $\cS_n.$
\end{lemma}

\begin{IEEEproof}
	We denote by $t_0$ the first time instance when all the nodes have failed at least once and note that $t_0$ is a stopping time for the
	sequence $(T_i)$. Under our model, the failures of a node are independent of other nodes and defined as a Poisson arrival process. For each node $v$, the probability that the node has not failed up to time $t$ is $e^{-\lambda t}$. Thus, we obtain 
	\[\Pr(t_0\leq t)=\prod_{i=1}^n \parenv{1-e^{-\lambda t}}=\parenv{1-e^{-\lambda t}}^n.\]
	This proves the finiteness claim. The uniform distribution of $\pi_{t_0}$ follows by symmetry.
\end{IEEEproof}

Since $t_0\le\infty$ a.s., the time $t'=t-t_0$ is well defined. Consider a continuous-time Markov chain $X(t')$ 
with the state space $\cS_n$ constructed 
as follows.  Let $l\in [n]$ and let $\tau_l=(l,n,n-1,\dots,l+1)$ be a permutation (in cycle notation) that moves entry $l$ to the last position, and shifts everything to the right of $l$ one step to the left. Then $P(\pi\to\sigma)=\frac 1n$ if and only if $\sigma=\tau_l\circ \pi$ for some $l$, and $ P(\pi\to\sigma)=0$ for all other pairs $\pi,\sigma.$

 Let $N(t')$ be the number of nodes that failed until time $t'$. This is a Poisson counting process with rate $n\lambda$, i.e., $N(t')\sim \text{Poi}(n\lambda)$. At time $t'=0$, $X(0)$ is chosen uniformly at random. For $t'\geq 0$ define $X(t')=\pi_{N(t')},$ where $\pi$ with an integer index is defined above before Example~\ref{example1}. Due to the memoryless property of the exponential distribution, we obtain that $X(t')$ is indeed a Markov chain. 

Next note that $X(t')$ is positive recurrent since the discrete-time chain on $\cS_n$ defined by the kernel $P$ is recurrent and the expected return time to a 
state in $X(t')$ is finite for any state in $\cS_n.$ For a positive recurrent continuous-time Markov chain, the limiting probability distribution $\mu$ is unique, 
exists almost surely, and is given by
\begin{align}
\label{eq:22}
\mu(\pi)=\lim_{\tau\to\infty} \frac{1}{\tau} \int_{0}^{\tau} \mathbbm{1}_{\pi}\parenv{X(t')}\mathrm{d}t'=\frac{1}{n\lambda \E\sparenv{(\pi\to \pi)}}
\end{align}
where $(\pi\to \pi)$ is the time to return to state $\pi$ starting from $\pi$ (See, for example, \cite[ p. 332]{Gal2013}). In our model, $\E\sparenv{(\pi\to \pi)}$ does not depend on $\pi$. In words, \eqref{eq:22} implies that for $t$ large enough, the time that the network spends in each state is almost the same. 
We use this fact next to find an upper bound for the capacity.

\begin{lemma}
	\label{lem:upper_bound_iccap}
	Let $(\cN,\obeta,\obbs,\ot,h)$ be a storage network, where $\obbs$ is a (random) sequence of failed nodes and $h$ is a weight function satisfying the constraints given by $\obeta$. Assume also that $h_j$ is a function of the last failed node, i.e., if $\obbs_{j}=v_{\ell}$ then $h_j=h_{v_{\ell}}$. Then almost surely 
	\begin{equation}\label{eq:upper}
	\ccap(\cN)\leq \frac{k'(2n-k'-1)}{2}\frac{1}{n}\sum_{i=1}^n \beta_i .
	\end{equation}
\end{lemma}

\begin{IEEEproof}
	The capacity of $\cN$ is equal to the minimum weight under $h$ of a cut between $\tilde{v}$ and $DC_t$ where $DC_t$ can connect to any set of $k'$ nodes from $\cA_t$. Assume that the set of weight functions is given by $\mathset{h_v}_{v\in V}$. Since there is a weight function for every node $v$, we will denote by $h_v(u)$ the weight that $h_v$ assigns to the edge $(u,CU)$. Let $t_{j_0}$ be the first time instance by which all the nodes have failed at least once.  According to Lemma \ref{lem:finitetime}, $t_{j_0}$ is almost surely finite. By \eqref{eq:22} we may assume that all permutations appear as associated permutations with equal probability. 
	
	Let $D:=\mathset{v_{i_1},\dots,v_{i_{k'}}}\subset V$ and assume that the associated permutation $\pi$ is such that \bcomment{ $\pi^{-1}(v_{i_1})\leq \pi^{-1}(v_{i_2})\leq \dots\leq \pi^{-1}(v_{i_{k'}})$.} Then the weight of the cut between $\tilde{v}$ and $D$ is at most \cite{DimGodWUWaiRam2010}
  \begin{equation}\label{eq:cth}
  C_t^h(D)\le \sum_{\ell=1}^{k'} \Big( \sum_{v\in V\setminus v_{i_{\ell}}}h_{v_{i_1}}(v)- \sum_{r=1}^{\ell-1}h_{v_{i_{\ell}}}(v_{i_r}) \Big).
  \end{equation}

	
	Since $\ccap(\cN)$ is the minimum weight value of a cut, we can bound it above by the average weight:
	\begin{align}
	\label{eq:ab1}
	\ccap(\cN)&\leq \frac{1}{n!\binom{n}{k'}} \sum_{\substack{D\subseteq V \\ |D|=k'}} \sum_{\pi_t\in\cS_n} C_t^h(D),
	\end{align}
	For the moment let us fix $D$ and consider how many times the term $h_v(u)$ appears on the right-hand side of \eqref{eq:ab1} as we substitute $C_t^h(D)$ from 
	\eqref{eq:cth} and evaluate the sum on $\pi_t.$ If both $u,v\in D$ then this term 
	appears for those $\pi_t$ in which $v$ appears after $u$ and does not 
	(is canceled in \eqref{eq:cth}) if $v$ precedes $u$. Thus, overall this term appears 
	$n!/2$ times. If $v\in D$ and $u\not\in D$ then no cancellations occur, and the 
	term $h_v(u)$ appears $n!$ times. Further, there are \bcomment{$\binom{n-2}{k'-2}$} choices 
	of $D$ for the first of these options and $\binom{n-2}{k'-1}$ for the second one 
	of them. Thus, for each pair of nodes $u,v\in V$ the term $h_u(v)$ appears in 
	\eqref{eq:ab1} 
    $$
    \binom{n-2}{k'-2}\frac{n!}{2}+\binom{n-2}{k'-1}n!=\binom{n-2}{k'-2}\frac{n!}{2}\frac{2n-k'-1}{k'-1}
    $$
    times. Substituting this into \eqref{eq:ab1} and performing cancellations, 
    we obtain that
	$$
	\ccap(\cN)\leq \frac{k'(2n-k'-1)}{2n(n-1)}\sum_{i=1}^n\sum_{j\neq i}h_{v_i}(v_j).
	$$
	Since $h$ is a weight function and since the nodes fail with equal probability, we 
	obtain that $\sum_{i\neq j}h_{v_i}(v_j)=(n-1)\beta_j, j=1,\dots,n$. Thus,  
	$$
	\sum_{i=1}^n\sum_{j\neq i}h_{v_i}(v_j)=(n-1)\sum_{j=1}^n \beta_j
	$$
	and the result follows.
\end{IEEEproof}

\begin{remark}\normalfont
	Lemma \ref{lem:upper_bound_iccap} holds also for $h_j$ that is a function of the current associated permutation, i.e., $h_j=h_{\pi_{t_j}}.$
\end{remark}
It is intuitively clear (and is confirmed by Lemmas \ref{lem:upper_bound_iccap} and \ref{lem:lower_bound_iccap}) that if $\beta_i=\beta_0$ for every $i\in [n]$, the fact that $\obbs$ is random does not affect the storage capacity, which implies that $\ccap(\cN)$ is equal to the minimum cut. Hence, in the case that $\beta_i=\beta_0$, both $\ccap(\cN)$ and its expected value are given in \cite{DimGodWUWaiRam2010}. 

\subsection{Discrete Time Evolution}\label{sec:discrete}
In this subsection we define a discrete-time storage network, which will enable us to simplify the analysis of 
the average cut in the information flow graph and of network capacity. 
A discrete-time storage network is a network $(\cN,\obeta,\os,\ot,h)$ with $\ot=(1,2,\dots)$. For such a network with weight function $h$, the average cut is defined as
\[C_{\rm avg}^h(\cN)=\limsup_{l\to\infty} \frac{1}{l}\sum_{t=1}^l C^h_t(\cN).\] 
For a discrete-time network $(\cN,\obeta,\os,\ot,h)$ we will sometimes omit the time notion $\ot$. Also, when a weight function is not specified we will omit the weight function notion and write $(\cN,\obeta,\os)$.
The following lemma shows that for a random discrete-time storage network with $h=\wt$, the limit superior in this definition is almost surely a limit. 
\begin{lemma}
	\label{lem:converge}
	Let $(\cN,\obeta,\obbs,\wt)$ be a random discrete-time storage network with $\obbs=(\bbs_i, i\ge 1)$ a sequence of independent RVs uniformly distributed on $[n].$ 
	Then 
	$$
	\E\sparenv{C_{\rm avg}^{\wt}(\cN)}{=}\lim_{l\to\infty} \frac{1}{l}\sum_{t=1}^l \E[C_t^{\wt}(\cN)].
	$$
\end{lemma}
\begin{IEEEproof}
	Let $t_0$ be the first time instance by which all the nodes have failed at least once. 
	Note that $t_0$ is a stopping time and each failed node is chosen uniformly and independently. 
	Referring to the Coupon collector's problem \cite[p.210]{GS01}, we obtain  that $\Pr(t_0\geq cn\log n)\leq n^{1-c}, $ for every $c\ge 1$. 
	Thus, $t_0$ is finite almost surely.
	
	By symmetry, $\pi_{t_0}$ is distributed uniformly on the set of all permutations. 
	Moreover, since $\bbs_i$ is chosen uniformly and independently, for $t\geq t_0$ we have that 
	$
	\Pr(\pi_t=\pi |
	\pi_0^{t-1})=\Pr(\pi_t=\pi | \pi_{t-1}),
	$
	so the sequence $\{\pi_t\}$ is a Markov chain, which is irreducible and aperiodic. Because of this, a limiting distribution $\mu$ exists, and is unique and
	positive.
	Hence, as $t$ 
	grows, $\Pr(\pi_t)\to \mu(\pi_t)$. Together with the fact that $C_t^{\wt}$ is {uniformly} bounded from above 
	for all $t,$ we obtain that the limit $\lim_{l\to \infty}  
	\frac{1}{l}\sum_{t=1}^l \E[C_t^{\wt}(\cN)]$ exists.
	
	Now define $X_t=\frac{1}{t}\sum_{i=1}^t C_i^{\wt}(\cN)$ and note that $X_t$ is a function of $\obbs$. Following the previous discussion, for almost every $\obbs$, the sequence $X_t$ converges. Since $X_t$ is non-negative and upper bounded for every $t$, by the dominated convergence theorem we have $\lim_{t\to\infty}\E[X_t]=\E[\lim_{t\to\infty}X_t]$ (the last limit exists a.s.), which is the desired result.
\end{IEEEproof}
Since $t_0$ is almost surely finite and since $\pi_t$ is an ergodic Markov chain, 
defining the initial state to be $\pi_0={\rm id}$ does not affect the expected capacity. Hence, from now on we assume $\pi_0={\rm id}$. 

The problem of finding the limiting distribution of our Markov chain on $\cS_n$ is similar to the 
classic question of the mixing time for the card shuffling problem called {\em Top in at random shuffle}. We use the following result from \cite[Thm.1]{AldDia86}.
\begin{theorem}	[{\sc Aldous and Diaconis}]\label{thm:diaconis}
	Consider a deck of $n$ cards. At time $t=1,2,\dots$ take the top card and insert it in the deck at a random position. 
	Let $Q_t$ denote the distribution after $t$ such shuffles and let $U$ be the uniform distribution on the set of all permutations 
	${\cS}_n.$ Then for all $c\geq0$ and $n\geq 2$, the total variation distance satisfies
	\begin{equation}\label{eq:AD}
	\|Q_{n\log n+cn}-U\|_{TV}\leq e^{-c}.
	\end{equation}
\end{theorem}
To connect this result to our problem, we note that choosing the next failed node uniformly at random corresponds to 
selecting a random card from the deck and putting in at the bottom.
The mixing time of this chain is stochastically equivalent to the mixing time of the {\em Top in at random shuffle}, and
we obtain the following lemma. 

\begin{lemma}
	\label{lem:uni}
	Let $\cN$ be a storage network with $|V|=n\geq 2$ nodes and let $\obbs$ be a random sequence of failed nodes. 
	Consider the sequence of associated permutations $(\pi_t, t\ge 0)$ where $\pi_0={\rm id}$. Then for any $c\geq 0$, $n\geq 2$ and 
	any  $\pi\in{\cS}_n$, 
	$$\Big|\Pr(\pi_{n\log n+cn}=\pi)- \frac{1}{n!}\Big|\leq e^{-c}.$$
\end{lemma}

\bcomment{\begin{IEEEproof} Let $T\ge 1$ be a time instant, and consider a realization $(\pi_t, t=0,2,\dots, T)$ of the Markov chain 
of failed nodes. Let $\tilde\pi_t, t\ge 0$ denote a realization of the card permutations in the top in at random shuffle.
Then 
   $$
   \Pr(\pi_T=\tau|\pi_0=\text{id})=\Pr(\tilde\pi_T=\text{id}|\tilde\pi_0=\tau)
   $$
for any $\tau\in\cS_n.$
Taking $T=n\log n +cn, c\geq0$ and using the definition of the total variation distance, we obtain the claimed result from \eqref{eq:AD}.
\end{IEEEproof}}

The next lemma, whose proof is given in Appendix~\ref{app:equalavg}, relates the values of average cut in the storage 
networks with continuous and discrete time.
\begin{lemma}\label{lem:equalavg}
	Let $(\cN_1,\obeta,\obbs,\ot,\wt)$ be a continuous-time storage network and let $(\cN_2,\obeta,\obbs,\wt)$ be a discrete-time storage network. Then
	$$C_{\rm avg}^{\wt}(\cN_1)\stackrel{a.s}{=} C_{\rm avg}^{\wt}(\cN_2).$$
\end{lemma}
\vspace*{.1in} As a result, we obtain the following statement which forms a basis of our subsequent derivations.
	\begin{theorem}
		\label{cor:first}
		Let $(\cN_1,\obeta,\obbs,\ot)$ be a continuous-time storage network. 
		Then ($(\mu_1\times\mu_2)$-a.s.) 
		  $$
		\ccap(\cN_1)\geq \frac{1}{n!}\sum_{\pi_t\in\cS_n} C_t^{\wt}(\pi_t).
		  $$
	\end{theorem}

\begin{IEEEproof}
	From Lemma \ref{lem:lower_bound_iccap} we have that for any realization $s$ such that every node fails infinitely often, 
	$\ccap(\cN_1)\geq C_{\rm avg}^{\wt}(\cN_1)$. According to Lemma \ref{lem:finitetime}, there exists a finite $t_0$ by which all the nodes have failed at least once and by \eqref{eq:22} the stationary distribution of the permutations is uniform. This implies that almost surely, all the nodes fail infinitely often. 
	According to Lemma \ref{lem:equalavg}, if $(\cN_2,\obeta,\obbs,\wt)$ is a discrete-time storage network, almost surely 
	$C_{\rm avg}^{\wt}(\cN_1)=C_{\rm avg}^{\wt}(\cN_2)$. From Lemmas \ref{lem:converge} and \ref{lem:uni}, $C_{\rm avg}^{\wt}(\cN_2)$ is almost surely a constant, which is equal to $\frac{1}{n!}\sum_{\pi_t\in\cS_n}C_t^{\wt}(\pi_t)$. Hence, almost surely, 
	$C_{\rm avg}^{\wt}(\cN_2)=\E\sparenv{C_{\rm avg}^{\wt}(\cN_2)}=\frac{1}{n!}\sum_{\pi_t\in\cS_n}C_t^{\wt}(\pi_t)$. 
Altogether these statements imply the claim of the theorem.
\end{IEEEproof}
 
\vspace*{.1in} From this point, unless stated otherwise, we restrict ourselves to discrete-time networks.

\section{The Fixed-Cost Model}\label{sec:static}

In this section, we define the fixed-cost storage model and derive lower bounds on the storage capacity. Suppose that the set of nodes
is $V=U\cup L,$ where $U=(v_1,\dots, v_{n_1})$ and ${L=(v_{n_1+1}\dots, v_{n_1+n_2})}$ are disjoint non-empty subsets. Suppose that
the repair bandwidth of the node $v_i$ is given by
     $$
       \beta_i=\begin{cases} \beta_1 &\text{if }v_i\in U\\\beta_2 &\text{if }v_i\in L\end{cases}.
       $$
where $\beta_1\ge \beta_2>0.$ 
 Let $C$ be the minimum cut of $\cN$ in the static case (i.e., the worst-case weight of the cut):
 \begin{align}
 \label{eq:SC}
 C\triangleq \min_{\pi\in\cS_n} \{C_t^{\wt}(\pi)\}=\min_{\substack{t\geq 0,\\ \os\in V^{\infty}}} \{C_t^{\wt}\}.
 \end{align}
 Let $a\triangleq k-n_1$ and let us assume that $a>0$ because otherwise the file reconstruction problem is trivially solved by contacting $k$ nodes in $U$. 
 The minimum cut is given by the following result from \cite{AkhKiaGha2010cost} (we cite it using our assumptions of $a>0$ and large $\alpha$).
\begin{lemma}
	\label{lem:mincutknown}
	Let $(\cN,\obeta,\os,\wt)$ be a fixed-cost storage network. 
	Then
 \begin{align}
C&= \frac{n_1(n_1-1)}{2}\beta_1 
+\parenv{n_2(n_1+a-1)-\frac{a(a+1)}{2}}\beta_2.\label{eq:C}
\end{align}
\end{lemma}

In this section we consider a dynamical equivalent of the above model, where the sequence of node failures $\obbs$ is random. 
   Note that if $n_2=1$ then $k=n$ which implies that no coding is used in the storage network, so we will assume that $n_2\ge 2.$ To avoid boundary 
cases, we will also assume that $n_1>1$ (the case of $n_1=1$ is not very interesting and can be handled using the same technique as below). 

Expression \eqref{eq:C} gives the size of the minimum cut in the static model of \cite{DimGodWUWaiRam2010} and it also gives a lower bound 
for the cut $C_t^{\wt}$ for all $t$ and $\os$ in the dynamical model. We shall now demonstrate by example that by controlling the transmission 
policy it is possible to increase the storage capacity of the $(\cN,\obeta,\obbs,h)$ network compared to \eqref{eq:C}.

The idea of the example is as follows. 
\bcomment{Assume that $s_j$ is a failed node that needs to be recovered. Recall that in the static case, every active node transmits a fixed number of symbols for the recovery of $s_j$, namely, the nodes in $U$ transmit $\beta_1$ symbols and the nodes in $L$ transmit $\beta_2$ symbols. In the dynamic case we can choose how many symbols each node transmits for the recovery of $s_j$ as long as the average constraint is satisfied. We change the number of symbols that node $v_j$ transmits for the recovery of $s_j$ depending on whether each of them is in $U$ or $L$ (the exact expressions are given in the example below). 
We then show that the average constraint is satisfied, and that this yields an increase of the capacity of the system.}

\begin{example}
	\label{ex:basic}
	\normalfont{Let $(\cN,\obeta,\obbs,h)$ be a storage network with $n=20,k'=13,$ $U=(v_1,\dots,v_{10})$, $L=(v_{11},\dots,v_{20}),$ and $\beta_1=2\beta_2$. 
		Assume that $\alpha$ is large enough (in this case taking $\alpha\geq 33.5\beta_2$ suffices). 
		By \eqref{eq:C}, the value of the minimum cut with $h=\wt$ is $214\beta_2,$ and thus the maximum file size that can be stored \bcomment{in the static case} is 
		$M=214\beta_2.$ 
		The task of node repair is accomplished by contacting $19$ nodes. 

Now we will show that, under the dynamic model, it is possible to increase the file size by using the weight function $h$ defined as follows. Suppose that at time $t$ (recall that time is discrete) a node $v\in U$ has failed, i.e. $\bbs_t=v$ where $v\in U$, and define 
		\[h_t(v_i)=\begin{cases}
		\beta_2 & v_i\in L \\
		\beta_1+\frac{1}{20}\beta_2 & v_i\in U\setminus v \\
		0 & v_i=v.
		\end{cases}\]
If $\bbs_t=v$ where $v\in L$, define 
		\[h_t(v_i)=\begin{cases}
		\beta_2 & v_i\in L\setminus v \\
		\beta_1-\frac{9}{200}\beta_2 & v_i\in U \\
		0 & v_i=v.
		\end{cases}\]
		A straightforward calculation of the minimum cut yields that 
		  \begin{equation}\label{eq:mcut}
		\min_{\pi\in\cS_{10}} \mathset{C_t^{h}(\pi)}=(214+2.25)\beta_2
		\end{equation}
		and it is obtained when $\pi=id$ and the active nodes selected are $D_t=(v_{1},v_{2},\dots,v_{13})$. This shows an increase over the 
		static case estimate \eqref{eq:C}.
		
		We now calculate the expected number of symbols a node transmits under $h$. Recall that in the random model, each node has the same probability of failure which in this case equals to $\frac{1}{20}$. Let $t_0$ denote the first time instance by which all the nodes have failed. For every $t\geq t_0$ we have that if
		$v_i\in U$ then 
		\[\E\sparenv{h_t(v_i)}=\frac{9}{20}(\beta_1+\frac{1}{20}\beta_2)+\frac{10}{20}(\beta_1-\frac{9}{200}\beta_2)<\beta_1\]
		and if $v_i\in L$ then 
		\[\E\sparenv{h_t(v_i)}=\frac{9}{20}\beta_2+\frac{10}{20}\beta_2<\beta_2.\]
		Therefore, the average amount of symbols each node transmits satisfies the constraints given by $\obeta$.
		
The above simple procedure is not optimal in terms of the file size $M$: As we show below, it is possible to 
construct a different transmission scheme which allows for storage of a larger-size file. Note also that the upper
bound \eqref{eq:upper} gives $\ccap(\cN)\le 235.5\beta_2,$ while the improvement of \eqref{eq:mcut} over \eqref{eq:C} is relatively minor.
	
	}
\end{example}

Example \ref{ex:basic} provides a procedure to construct the weight function $h$ such that the maximum file size can be increased. 
Below we generalize this idea and also explore other ways of using time evolution to increase the storage capacity of a fixed-cost network
	
\subsection{A protocol to increase capacity}
In this section we construct a weight function that increases the storage capacity and analyze the increase. The next theorem states the increase explicitly. 
In order to state the theorem, we need the following assumptions. For $\varepsilon_1>0$, assume that
\begin{align}
\beta_1-\beta_2&\geq \frac{n(n_1-1)}{n_2}\varepsilon_1. \label{as:1}  
\end{align}
Note that $\frac{n(n_1-1)}{n_2}\geq 1$ and that this assumption is satisfied in Example \ref{ex:basic} above.

We now prove the following theorem which quantifies the increase of the average storage capacity over the static case.
\begin{theorem}
	\label{thm:main12}
	Let $(\cN,\obeta,\obbs)$ be a fixed-cost storage network. For any $\varepsilon_1\geq 0$ such that assumption \eqref{as:1} is satisfied, the storage capacity is bounded below by  
	$$
	\ccap(\cN)\stackrel{\rm a.s.}{\geq} C+\frac{n_1(n_1-1)}{2}\varepsilon_1
	$$
	where $C$ is the static storage capacity given in \eqref{eq:C}.
\end{theorem}

To prove the theorem we define a weight function along the lines of Example \ref{ex:basic}. 
Let us put $h_t(v_i)=h_U(v_i)$ if $\os_t\in U$ and $h_t(v_i)=h_L(v_i)$ if $\os_t\in L$ where
  \begin{equation}\label{eq:hU}
  h_U(v_i)=\begin{cases}
\beta_1+\varepsilon_1 & v_i\in U\setminus \os_j \\
\beta_2 & v_i\in L \\
0 & v_i=\os_j,
\end{cases}
  \end{equation}
and 
  \begin{equation}\label{eq:hL}
  h_L(v_i)=\begin{cases}
\beta_1-\frac{n_1-1}{n_2}\varepsilon_1 & v_i\in U \\
\beta_2 & v_i\in L\setminus \os_j \\
0 & v_i=\os_j
\end{cases}
  \end{equation}
and $0\leq \varepsilon_1\leq \beta_1$.

We now show that the weight function $h$ satisfies the constraints given by $\beta$.
\begin{lemma}
	\label{lem:constsat}
	Let $(\cN,\obeta,\obbs,h)$ be a fixed-cost storage network with $h$ as defined above. Then $h$ satisfies the average constraints given by $\obeta$.
\end{lemma}

\begin{IEEEproof}
	Fix a node $v_i\in U$ and for each time instance $t$, let us calculate the expected number of symbols $v_i$ transmits. 
	Recall that $v_{\pi_t(n)}$ denotes the node that failed at time $t$. Recall that the failures of the nodes are uniformly distributed, so we obtain 
	  $$\Pr(v_{\pi_t(n)}=v_j)=\begin{cases}
	\frac{1}{n} & \text{ if } j=i \\
	\frac{n_1-1}{n} & \text{ if } j\in [n_1]\setminus i \\
	\frac{n_2}{n} & \text{ otherwise}.
	\end{cases} 
	  $$
	Hence, the expected number of symbols that the node $v_i$ transmits is 
	\begin{align*}
	 &\frac{n_1-1}{n}h_U(v_i)+\frac{n_2}{n}h_L(v_i) 
	 =\frac{n_1-1}{n}(\beta_1+\varepsilon_1)+\frac{n_2}{n}\parenv{\beta_1-\frac{n_1-1}{n_2}\varepsilon_1}
	 < \beta_1.
	 \end{align*}
	If $v_i\in L$ we have 
	\[\Pr(v_{\pi_t(n)}=v_j)=\begin{cases}
	\frac{1}{n} & \text{ if } j=i \\
	\frac{n_1}{n} & \text{ if } j\in [n_1]\\
	\frac{n_2-1}{n} & \text{ otherwise}.
	\end{cases} \]
In this case the expected number of trnasmitted symbols equals
	\begin{align*}
	\frac{n_1}{n}h_U(v_i)+\frac{n_2-1}{n}h_L(v_i)
	&=\frac{n_1}{n}\beta_2+\frac{n_2-1}{n}\beta_2
	< \beta_2.
	\end{align*}
	Thus, on average the number of symbols is within the allotted bandwidth.
\end{IEEEproof}

The next two lemmas are used in the proof of Theorem \ref{thm:main12} in order to estimate the minimum cut. The first lemma shows that the minimum cut for any permutation $\pi_t, t\geq t_0$ is obtained when $D_t\supseteq U$. The second lemma shows that the minimum cut is obtained for $\pi_t=id$.
\begin{lemma}
	\label{lem:opchoose}
	Let $(\cN,\obeta,\os,h)$ be a network with $h$ as defined above. 
	If assumption \eqref{as:1} is satisfied, then for $t>t_0$, the value $C_t^{h}(\cN)$ is attained when $D_t\supseteq U$.
\end{lemma}

\begin{figure}[t!]
    \centering
    \includegraphics[width=.8\linewidth]{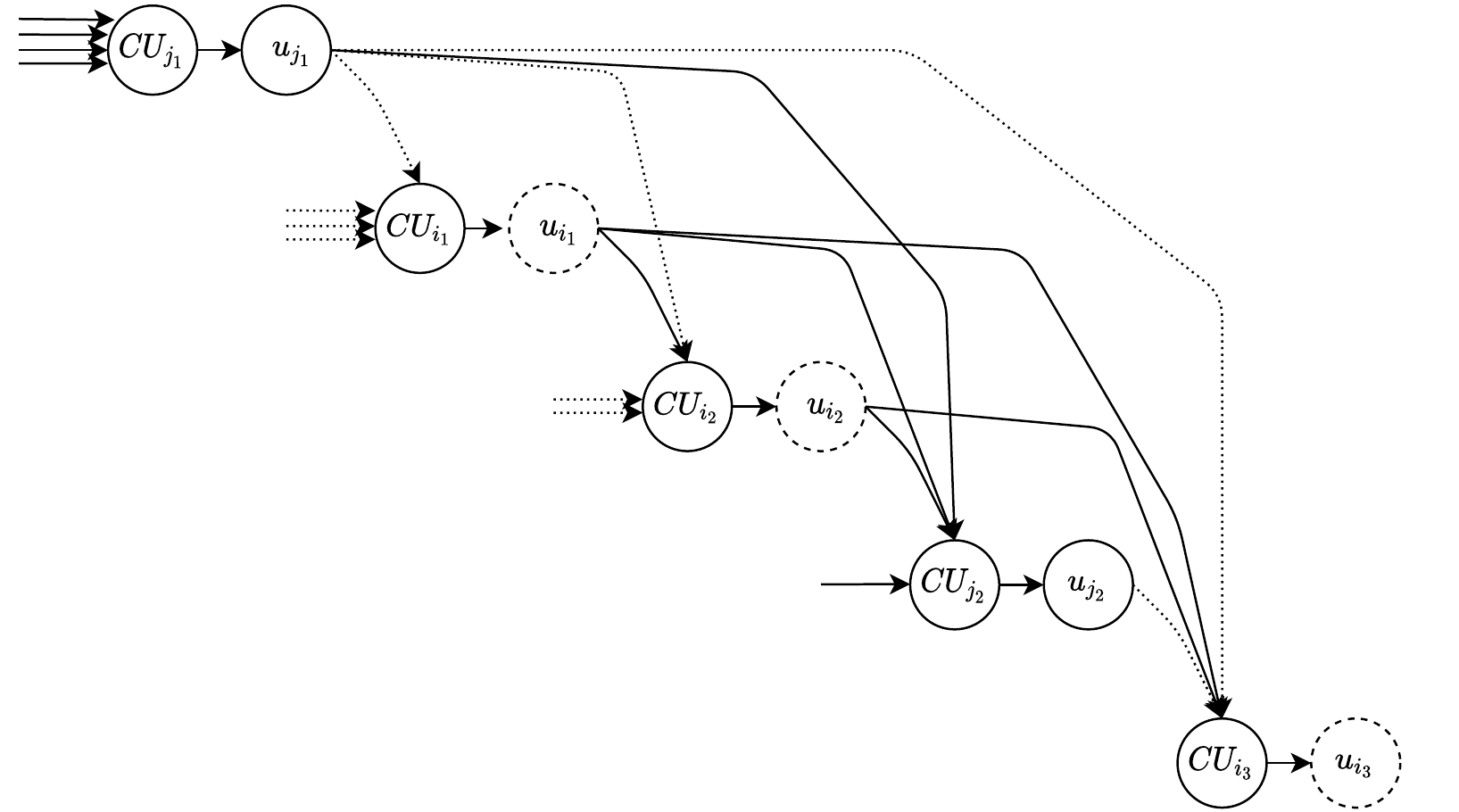}
    \caption{\bcomment{An illustration of the procedure in the proof of Lemma \ref{lem:opchoose} with $\kappa=4,$ where the 3 dashed nodes were selected by the
$DC_t,$ and we are to select the next node out of $u_{j_1},u_{j_2}$. The current cost of the procedure equals the sum of the weights of the dotted edges. Suppose that node $u_{j_2}$ is selected next. Then the weight of the edge from 
$u_{j_1}$ to $CU_{j_2}$ (which equals $h_{j_1}(u_{j_2})$) and the weight of the solid in-edge of $CU_{j_2}$ will be added to the cost, while the weight of the edge from $u_{j_2}$ to $CU_{i_3}$ (which equals $h_{j_2}(u_{i_3})$) will be deducted. Otherwise, if node $u_{j_1}$ is selected, 
the weights of the solid in-edges of $CU_{j_1}$ will be added and the weights of all the dotted out-edges of $u_{j_1}$ will be deducted.}}
    \label{fig:2}
\end{figure}

\begin{IEEEproof}
	We formulate our question as a dynamic programming problem and provide an optimal policy for node selection. 
	Assume that $\pi_t$ is a fixed permutation that represents the order of the last $n$ failed nodes. 
	We will consider the information flow graph $\cX_t$ and show that the cut is minimized when all the nodes from $U$ are selected. 
	
	Consider a $k'$-step procedure which in each step selects one node from $\cA_t.$
	Each step entails a cost \bcomment{as explained next}. 
	Let $t'\leq t$ and assume that node $v_{i_{t'}}^{t'}\in\cA_t$ was selected. 
	The cost is defined as the added weight values of the in-edges of $CU_{t'}$ that are not out-edges of previously selected nodes. 
	Our goal is to choose $k'$ nodes that minimize the total cost and hence minimize the cut between $\parenv{\bigcup_{j=-1}^{t-1}\cA_j}\setminus \cA_t$ and $DC_t$. 
	
	In order to simplify notation, we write $\pi_t=(u_1,u_2,\dots,u_n)$, i.e., $u_l=v_{\pi_t(l)}$ is the storage node that appears in the $l$th position in $\pi_t$. 
	Moreover, with a slight abuse of notation, if $u_j$ failed at time $t'$ we will write $h_{j}(u_i)$ instead of $h_{t'}(u_i)$ \bcomment{to denote the number of symbols that node $u_i$ transmits for the recovery of $u_j$}.
	For $\kappa\le k'$ consider the sub-problem in step $\kappa-1$, where the $DC_t$ has already chosen 
	$\kappa-1$ nodes $(u_{i_1},\dots, u_{i_{\kappa-1}})$ and we are to choose the \bcomment{$\kappa$th }node. 
	Assume that the chosen nodes are 
	ordered according to their appearance in the permutation, i.e., $i_1\leq i_2\leq\dots\leq i_{\kappa-1}$. 
	Let  $u_{j_1},\dots,u_{j_m}\in U$ be nodes that were not 
	selected up to step $\kappa-1$, i.e.,  
	\[\{u_{j_1},\dots, u_{j_{m}}\}\cap \{u_{i_1},\dots, u_{i_{\kappa-1}}\}=\emptyset,\] 
	and assume also that $j_1\leq j_2\leq \dots \leq j_m$. 
	We refer to Figure \ref{fig:2} for an illustration.
	We show that choosing $u_{j_1}$ accounts for the minimum cut. 
	First, we claim that choosing $u_{j_1}$ minimizes the cut over all other nodes from $U$. Denote by $C_{\kappa-1}$ the total cost (or the cut) in step $\kappa-1.$ 
Fix $2\leq \ell\in[m]$ and note that since $j_1\leq j_{\ell},$ we can write 
	\begin{align*}
	i_1\leq \dots \leq i_{r_1}\leq j_1\leq i_{r_1+1}
	\leq \dots\leq i_{r_{\ell}}\leq j_{\ell}\leq i_{\ell+1} \leq \dots,
	\end{align*}
	where the set of indices $\{i_1,\dots,i_r\}$ can be empty.
Let $C(j_1)$ be the value of the cut once we add $u_{j_1}$ in the $\kappa$th step. The change from $C_{\kappa-1}$ is formed of the following
components. First, we add the values of all the edges from $U\backslash\{u_{j_1}\}$ to $u_{j_1}$ and from $L$ to $u_{j_1}$, accounting
for $(n_1-1)(\beta_1+\varepsilon_1)+n_2\beta_2$ symbols. Further, we remove the values of all the edges from the nodes $u_{i_1},\dots,u_{r_1}$
to $u_{j_1}$ and all the edges from $u_{j_1}$ to $u_{r_1+1},\dots u_{\kappa-1}.$ Overall we obtain
	   \begin{align}
	       \label{eq:cj1}
C(j_1)&=C_{\kappa-1}+(n_1-1)(\beta_1+\varepsilon_1)+n_2\beta_2
	-\sum_{q=1}^{r_1}h_{j_1}(u_{i_q})-\sum_{q=r_1+1}^{\kappa-1}h_{i_q}(u_{j_1}).
	   \end{align}
Similarly, let $C_{j_\ell}$ be the value of $C_{\kappa}$ if in step $\kappa$ we select the node $u_{j_\ell},\ell\ge 2.$ Following the same argument as in
\eqref{eq:cj1}, we obtain
        \begin{align*}
C(j_{\ell})&=C_{\kappa-1}+(n_1-1)(\beta_1+\varepsilon_1)+n_2\beta_2
	-\sum_{q=1}^{r_{\ell}}h_{j_{\ell}}(u_{i_q})- \sum_{q=r_{\ell}+1}^{\kappa-1}h_{i_q}(u_{j_{\ell}}).
    	\end{align*}
Since $h_{j_{\ell}}(u_{i})=h_{j_1}(u_{i})$ and $h_{i}(u_{j_1})=h_{i}(u_{j_{\ell}})$ for all $i\in [n]$, we have
	$$
C(j_1)-C(j_{\ell})= \sum_{q=r_1+1}^{r_{\ell}} \parenv{h_{j_{\ell}}(u_{i_q}) -h_{i_q}(u_{j_1})}.
    $$
For $u_{i_q}\in U$, we obtain 
     $$
h_{j_{\ell}}(u_{i_q}) -h_{i_q}(u_{j_1})=\beta_1+\varepsilon_1-(\beta_1+\varepsilon_1)=0.
    $$
For $u_{i_q}\in L$, we obtain 
	$$
h_{j_{\ell}}(u_{i_q}) -h_{i_q}(u_{j_1})=\beta_2-\Big(\beta_1-\frac{n_1-1}{n_2}\varepsilon_1\Big)
     $$
which is nonpositive by assumption \eqref{as:1}. 
Therefore, 
	\[C(j_1)-C(j_{\ell}) \leq 0.\]
	
Now we show that $u_{j_1}$ minimizes the cut over a selection of any node $u_{j_{\ell}}$ from $L$. We divide the argument into $2$ cases:
	\begin{enumerate}
		\item Assume that $j_{\ell}< j_1$. Denote by $(i_1,\dots,i_{r_{\ell}},j_{\ell},i_{r_{\ell}+1},\dots,i_{r_1},j_1,\dots)$ the indices of the selected nodes and let $C(j_{\ell}),C(j_1)$ be the cut values if we choose $u_{j_{\ell}},u_{j_1}$, respectively. We have
		\begin{align*}
C(j_{\ell})&=C_{\kappa-1}+n_1\Big(\beta_1-\frac{n_1-1}{n_2}\varepsilon_1\Big)
+(n_2-1)\beta_2- \sum_{q=1}^{r_{\ell}}h_{j_{\ell}}(u_{i_q})
-\sum_{q=r_{\ell}+1}^{\kappa-1}h_{i_q}(u_{j_{\ell}}).
		\end{align*}
On account of \eqref{eq:cj1} and \eqref{as:1} we now obtain
		\begin{align} 
		C(j_1)-C(j_{\ell})&=-(\beta_1-\beta_2) +\frac{n(n_1-1)}{n_2}\varepsilon_1 
		 +\sum_{q=1}^{r_{\ell}} h_{j_{\ell}}(u_{i_q}) -\sum_{q=1}^{r_{1}} h_{j_1}(u_{i_q}) \notag\\
				&\hspace*{.3in}+ \sum_{q=r_{\ell}+1}^{\kappa-1} h_{i_q}(u_{j_{\ell}}) - \sum_{q=r_{1}+1}^{\kappa-1} h_{i_q}(u_{j_{1}})
				\notag\\
		&\hspace*{-.5in}\le \sum_{q=1}^{r_{\ell}} \parenv{h_{j_{\ell}}(u_{i_q})-h_{j_1}(u_{i_q})} -\sum_{q=r_{\ell}+1}^{r_{1}} h_{j_1}(u_{i_q})
		+\sum_{q=r_{\ell}+1}^{r_1} h_{i_q}(u_{j_{\ell}})+\sum_{q=r_{1}+1}^{\kappa-1} \parenv{h_{i_q}(u_{j_{\ell}})- h_{i_q}(u_{j_{1}})}
		\label{eq:diff13}
		\end{align} 
Our goal is to show that the right-hand side of \eqref{eq:diff13} is nonpositive.
Let $1\le q\le r_\ell.$ For $u_{i_q}\in U$ we have 
		\begin{align*}
		&h_{j_{\ell}}(u_{i_q})-h_{j_1}(u_{i_q})
		=\beta_1-\frac{n_1-1}{n_2}\varepsilon_1 -(\beta_1+\varepsilon_1)
		\leq 0
		\end{align*}
		and for $u_{i_q}\in L$ we have 
		\[h_{j_{\ell}}(u_{i_q})-h_{j_1}(u_{i_q})=\beta_2-\beta_2= 0.\]
Now let $r_{\ell+1}\le q\le \kappa-1.$ 		
For $u_{i_q}\in U$ we have 
		$$
h_{i_q}(u_{j_{\ell}})- h_{i_q}(u_{j_{1}})= \beta_2-(\beta_1+\varepsilon_1)
       $$
and for $u_{i_q}\in L$ we have 
		\[h_{i_q}(u_{j_{\ell}})- h_{i_q}(u_{j_{1}})= \beta_2-(\beta_1-\frac{n_1-1}{n_2}\varepsilon_1),\]
both of which are non-positive by assumption \eqref{as:1}.
		
The remaining terms in \eqref{eq:diff13} contribute $\sum_{q=r_{\ell}+1}^{r_1} \parenv{h_{i_q}(u_{j_{\ell}})-h_{j_1}(u_{i_q})}$
to the value of the cut. As before, for $u_{i_q}\in U$ we have 
		$$
		h_{i_q}(u_{j_{\ell}})-h_{j_1}(u_{i_q})=\beta_2- (\beta_1+\varepsilon_1)\le 0
		$$
by \eqref{as:1}, and for $u_{i_q}\in L$ we have 
		$$
		h_{i_q}(u_{j_{\ell}})-h_{j_1}(u_{i_q})=\beta_2- \beta_2=0.
		$$
Thus, $C(j_1)-C(j_{\ell})\leq 0$.

\vspace*{.1in}		\item Assume that $j_{\ell}> j_1$. This case is symmetric to the case $j_{\ell}<j_1$ and the analysis is similar. 
	\end{enumerate}

\vspace*{.1in}
By the principle of optimality in dynamic programming, which states that every optimal policy consists only of optimal sub-policies  \cite[Ch. 1.3]{Ber2005}, we now conclude that the minimum cut is formed by first taking all the nodes from $U$ and then take the remaining nodes from $L$.
\end{IEEEproof}

\begin{remark}
	\label{rem:opchoose}
	\normalfont Suppose that in forming the cut, we have added all the nodes from $U$, and there are $a$ more nodes (from $L$) to select. To minimize
the value of the cut, these nodes should be taken to be the $a$ most recently failed nodes from $L$. This is because 
choosing the most recently failed node $v_{\pi(n)}$ assures that as few as possible of the previously selected nodes contain information from 
$v_{\pi(n)}$.

To justify this formally, consider the proof of Lemma \ref{lem:opchoose}. Indeed, if $u_{j_1},u_{j_{\ell}}\in L$ with $j_1<j_{\ell}$, 
	then
    $$
    C(j_1)-C(j_{\ell})=\sum_{q=r_1+1}^{r_{\ell}} h_{j_{\ell}}(u_{i_q})-h_{i_q}(u_{j_1})
    $$
	which is non-negative by assumption \eqref{as:1}.
\end{remark}

Before stating the second lemma, we need the following notation.
Let $\pi\in\cS_n$ and let $D\subset V, |D|=k'.$ 
For a node $v_j\in D$, denote by $f_\pi(v_j)$ the number of nodes in $D\cap L$
that appear before $v_j$ in $\pi$, i.e., 
 \bcomment{  $$
f_\pi(v):= \sum_{i=1}^n\mathbbm{1}_{D\cap L}(v_i)\cdot \mathbbm{1}_{[1,\pi^{-1}(v_j)]}(\pi^{-1}(v_i)).
   $$ }
Let $T_j,j=1,\dots,n-1$ be an {\em adjacent transposition} of $\pi,$ i.e., $T_j\circ \pi$ exchanges $\pi(j)$ and $\pi(j+1).$

\begin{lemma}
	\label{lem:cap1}
	Let $(\cN,\obeta,\obbs,h)$ be a fixed-cost storage network with $h$ as defined above. Let $\pi_t$ be a permutation obtained at time $t>t_0$ and let $D_t$ be a set of $k'$ active nodes selected by the $DC_t$. If assumption \eqref{as:1} is satisfied, then
	\begin{align*}
	C_t^h(\pi_t)&=C_t^h(id)+\sum_{v\in D_t\cap U} f_{\pi_t}(v)(\beta_1-\beta_2)
	-\sum_{v\in D_t\cap U} f_{\pi_t}(v)\frac{n_1-1}{n_2}\varepsilon_1,
	\end{align*}
	where $C_t^h(id)$ is the minimum cut for $\pi_t=id$.
\end{lemma}

\begin{IEEEproof}
	 Recall that by assumption \eqref{as:1}, $\beta_1-\beta_2-\frac{n_1-1}{n_2}\varepsilon_1\geq 0$. 
	 We start with showing that for every permutation $\pi_t\in \cS_n$ and for any $j\in [n]$, 
	\begin{align*}
	C_t(T_j(&\pi_t))\in \Bigg\{C_t(\pi_t)+(\beta_1-\beta_2)+\frac{n_1-1}{n_2}\varepsilon_1,
	C_t(\pi_t), C_t(\pi_t)-(\beta_1-\beta_2)-\frac{n_1-1}{n_2}\varepsilon_1\Bigg\}.
	\end{align*}

	First, observe that if $\pi_t$ and $\sigma_t$ are two permutations such that 
	\bcomment{$$
	\{\pi_t^{-1}(v_i) ~:~ i\in [n_1]\}=\{\sigma_t^{-1}(v_i) ~:~ i\in [n_1]\},
	$$}
	i.e., if the nodes from $U$ occupy the same positions in $\pi_t$ as in $\sigma_t$, 
	then $C_t(\pi_t)=C_t(\sigma_t)$. 
	
Let $\pi_t\in \cS_n$ and let $D_t$ denote the $k'$ storage nodes selected. Assume that $v_{\pi_t(j)}\in U$,  
	$v_{\pi_t(j+1)}\in L$ and that $\{v_{\pi_t(j)},v_{\pi_t(j+1)}\}\subseteq D_t$ for some $j\in [n-1]$.
It is easy to see that 
	\[C_t^h(T_j(\pi_t))=C_t^h(\pi_t)+(\beta_1-\beta_2)-\frac{n_1-1}{n_2}\varepsilon_1.\]
	On the other hand, if $v_{\pi_t(j)}\in L$, $v_{\pi_t(j+1)}\in U$ and $\mathset{v_{\pi_t(j)},v_{\pi_t(j+1)}}\subseteq D_t$, then 
      $$
 C_t^h(T_j(\pi_t))=C_t^h(\pi_t)-(\beta_1-\beta_2)+\frac{n_1-1}{n_2}\varepsilon_1.
       $$

	Recall that according to Lemma \ref{lem:opchoose}, the set $D_t$ which yields the minimum cut contains $U$. 
	Hence, for $id$, the minimum cut is given by $C_t(id)$ and is obtained by selecting the first $k'$ nodes, $D_t=\mathset{v_i ~:~ i\in [k']}$. Moreover, 
	every permutation $\pi_t\in\cS_n$ can be obtained from $id$ by repeated applications of $T$, such that at each application, the size of the minimum cut is not decreased.

\end{IEEEproof}

Note that Lemma \ref{lem:mincutknown} is an immediate corollary of Lemmas \ref{lem:opchoose} and \ref{lem:cap1}. 
Indeed, taking the weight function $h=\wt$ implies that 
$\varepsilon_1=0$ which satisfies assumption \eqref{as:1}. Hence, the minimum cut is obtained 
when $\pi_t=id$ and  $D_t\supseteq  U$, and is equal to $C$.

\vspace*{.1in}
Let us prove Theorem \ref{thm:main12}.
\begin{IEEEproof}[Proof of Theorem \ref{thm:main12}]
	From Lemma \ref{lem:opchoose} we obtain that there exists $\varepsilon_1>0$ such that assumption \eqref{as:1} is satisfied and such that at each time $t$, the selection $D_t$ that minimizes the cut, contains $U$.  Lemma \ref{lem:cap1} implies that the minimum cut is obtained
	for $\pi_t=id$. Taking $\pi_t=id$ and $D_t=\mathset{v_1,\dots,v_{k'}}$,  it is straightforward to check that 
	\begin{align}
	\label{eq:inc1}
	C_t^{h}(D_t)&= \sum_{j=1}^{n_1-1}j(\beta_1+\varepsilon_1)+n_1n_2\beta_2
	+\sum_{j=1}^{a}(n_2-j)\beta_2
	=C_t^{\wt}(\pi_t)+\frac{n_1(n_1-1)}{2}\varepsilon_1
	\end{align}
which together with Theorem \ref{cor:first} concludes the proof.
\end{IEEEproof}

\begin{remark}\normalfont
	The function $h$ can be defined with an additional parameter $0\leq \varepsilon_2\leq \beta_2$ such that when a node $v_i\in U $ $(v_i\in L)$ fails, nodes from $L$ transmit $\beta_2-\varepsilon_2$ (resp., $\beta_2+\varepsilon_2$) symbols instead of $\beta_2$ symbols. This change may increase the storage capacity even more, but requires additional assumptions on the parameters and can be developed along the same ideas.
\end{remark}

In conclusion, we have shown that the maximum file size that can be stored in a dynamical fixed-cost storage network is always greater
than its static counterpart. While it is always possible to choose $\varepsilon_1$ so that \eqref{as:1} holds true 
(e.g., $\varepsilon_1=\frac{n_2}{n(n_1-1)}(\beta_1-\beta_2)$), the capacity increase is relatively small because the allowable values of 
$\varepsilon_1$ are small as a proportion of $\beta_1-\beta_2.$ In the next section we take an alternative approach to bounding the average
capacity.

\subsection{The average min-cut bound on $\ccap$}\label{sec:general}

We consider the same storage model as in the previous section and prove the following result.

\begin{theorem}
	\label{thm:main1}
	Let $(\cN,\obeta,\obbs,\wt)$ be a fixed-cost storage network. Then almost surely, 
	\begin{equation}\label{eq:f}
		\textstyle{\ccap(\cN)\geq C+ \frac{\beta_1-\beta_2}2\frac{an_1}{n}\big(a+1+\frac{n_1-1}{n-1}(a-1)\big)}.
	\end{equation}
\end{theorem}

We will need the following two lemmas.

\begin{lemma}
	\label{lem:prob1}
	Let $(\cN,\obeta,\obbs)$ be a storage network, let $0\leq \ell\leq \min(n_1,a)$ and denote by $P_t^{\ell}$ the probability that $\pi_t$ contains $\ell$ nodes from $U$ in the last $a=k'-n_1$ positions. 
	As $t\to \infty$, 
	$$
	P_t^{\ell}\to\binom{n_1}{\ell}\binom{n_2}{a-\ell}\binom{n}{a}^{-1}.
	$$
\end{lemma} 

\begin{IEEEproof}
	Assume that $\pi_t$ is distributed uniformly over $\cS_n$. We have 
	$P_t^{\ell}=\binom{n_1}{\ell}\binom{n_2}{a-\ell}\binom{n}{a}^{-1}$. 
	By Lemma \ref{lem:uni}, the distribution of $\pi_t$ converges to the uniform distribution exponentially fast (after a certain time, the TV distance decreases by a factor of $1/e$ every $n$ time units). By the definition of the total variation distance, 
	for every $\ell$ 
	$$
	\abs{P_{t}^{\ell}-\binom{n_1}{\ell}\binom{n_2}{a-\ell}\binom{n}{a}^{-1}}\leq e^{\log n -\frac{t}{n}}
	$$
which implies the lemma.
\end{IEEEproof}

For the next lemma we need the following notation. Let $\cS_n^{\ell}$ be the set of all permutations over $[n]$ with exactly $\ell$ numbers from $U$ in the last $a$ positions, i.e.,
\begin{align*}
\cS_n^{\ell}\triangleq 
\mathset{\pi\in\cS_n : \abs{\mathset{\pi(n-a+1),\dots,\pi(n)}\cap [n_1]}=\ell}.
\end{align*}
Given $\pi=(i_1,\dots,i_{n-a},i_{n-a+1},\dots,i_n)\in \cS_n,$ let $\pi^c:=(i_1,\dots,i_{n-a},i_{n},\dots,i_{n-a+1})$.
\begin{lemma}
	\label{lem:perm}
	Let $(\cN,\obeta,\obbs,\wt)$ be a fixed-cost storage network.
	Let $\pi_t$ be the permutation at time $t$ and for every $\ell$, define $\mu_{\ell}(\pi_t):=\Pr(\pi_t | \cS_n^{\ell})$.
	Then, 
	$$\lim_{t\to\infty}\E_{\mu_{\ell}}[C_t^{\wt}(\cN)]\geq C+\frac{1}{2}\ell(a+\ell)(\beta_1-\beta_2)$$
	where $C$ is given in Lemma \ref{lem:mincutknown}.
\end{lemma}

\begin{IEEEproof}
	As above, let $t_0$ be time by which all the nodes have failed at least once, and recall that $P(t_0<\infty)=1.$ 
	Therefore, $\pi_t$ (and hence, $\mu_{\ell}$) is well defined almost surely. From Lemma \ref{lem:uni}, we obtain that for every $\epsilon>0$, there exists $t_{\epsilon}>t_0$ large enough such that $\abs{\mu_{\ell}(\pi_t)-\frac{1}{|\cS_n^{\ell}|}}\leq \epsilon$ and therefore the limit exists almost surely. 
	
	For $t>t_{\epsilon}$ consider
    $$
    \sum_{\pi_t\in\cS_n^{\ell}}\Pr(\pi_t | \cS_n^{\ell}) C_t(\pi_t)\geq \Big(\frac{1}{|\cS_n^{\ell}|-\epsilon}\Big)\sum_{\pi_t\in\cS_n^{\ell}} C_t(\pi_t)\geq 
    \frac{1}{|\cS_n^{\ell}|}\sum_{\pi_t\in\cS_n^{\ell}} C_t(\pi_t) -\epsilon R,
    $$
    where $R=\max_{\pi_t\in\cS_n} C_t(\pi_t)$. 
    To bound this sum below we fix the last $a$ entries of the permutation. Since for $h=\wt$ 
(i.e., $\varepsilon_1=0$), assumption \eqref{as:1} is satisfied. Thus we can use Lemma \ref{lem:opchoose}, 
according to which $C_t(\pi_t)$ is minimized if $n_1-\ell$ entries from $U$ appear in the first $n_1-\ell$ positions, followed by $n_2-a+\ell$ entries from $L$ 
    (in any order). Fix the first $n-a$ entries. Again according to 
    Lemma \ref{lem:opchoose}, the minimum cut will be obtained when all the 
    $\ell$ nodes from $U$ are in positions $n-a+1,n-a+2,\dots,n-a+\ell,$ and 
    according to Lemma \ref{lem:cap1} it is equal 
    to $C_{\min}:=C+\ell^2(\beta_1-\beta_2)$. Also, the maximum cut will be 
    obtained when all the $\ell$ nodes from $U$ are located in the last positions. 
    This yields $C_{\max}:=C+\ell a(\beta_1-\beta_2)$. 
	
	Let $\pi_t\in\cS_n^{\ell}$ be any permutation with $v_{\pi_t(i)}\in U$ for $i\in \{1,\dots,n_1-\ell\}.$  We claim that
     \begin{align}
     \label{eq:equal1}
     C_t(\pi_t)+C_t(\pi_t^c)&=2C+\ell(a+\ell)(\beta_1-\beta_2)
	=C_{\min}+C_{\max}.
	\end{align}
	Indeed, assume $\pi_t=\pi$ and let $D$ be a selection of $k$ active nodes that minimizes the cut. By Lemma \ref{lem:opchoose} if there is at least one node from $U$ in 
	the last $a$ places, the minimum cut will be obtained by selecting the last $a$ places as a part of $D$. Moreover, if $v_i\in U$ with \bcomment{$\pi^{-1}(v_i)=n-a+m$} for 
some $m\in [a]$, and $f_\pi(v_i)=b$ then $\abs{\mathset{v_{\pi(1)},\dots,v_{\pi(n-a+m)}}\cap (D\cap L)}=b$. 
Together with the fact that $\abs{D\cap L}=a$, this implies that 
$\abs{\{v_{\pi(n-a+1)},\dots,v_{\pi(n-a+m)}\}\cap  (D\cap L)}=b-\ell$. For $\pi^c$, we obtain that \bcomment{$(\pi^c)^{-1}(v_i)=n-m+1$} and 
$\abs{\{v_{\pi^c(n-m+1)},\dots,v_{\pi^c(n)}\}\cap L}=b-\ell$ which means that $\abs{\{v_{\pi^c(1)},\dots,v_{\pi^c(n-m+1)}\}\cap (D\cap L)}=a-(b-\ell)$. 

By Lemma \ref{lem:cap1} we have 
    \begin{align*}
C_t(\pi)&= C+\sum_{v\in D\cap U}f_{\pi}(v)(\beta_1-\beta_2) 
\geq C+\sum_{\substack{v\in D\cap U\\ \pi^{-1}(v)\in\mathset{n-a+1,\dots, n}}}f_{\pi}(v)(\beta_1-\beta_2).
    \end{align*}
    For $\pi^c$ we obtain 
     \begin{align*}
     C_t(\pi^c)
     &\geq C\hspace*{.1in}+\hspace*{-.1in}\sum_{\substack{v\in D\cap U\\ (\pi^c)^{-1}(v)\in\mathset{n-a+1,\dots, n}}}f_{\pi^c}(v)(\beta_1-\beta_2) \\
	 &= C \hspace*{.1in}+\hspace*{-.1in}\sum_{\substack{v\in D\cap U\\ 
	 		(\pi^c)^{-1}(v)\in\mathset{n-a+1,\dots, n}}} \hspace{-1cm}(a-(f_{\pi}(v)-\ell))(\beta_1-\beta_2).
 	\end{align*}
	This implies that 
	\[C_t(\pi)+C_t(\pi^c)\geq 2C+\ell(a+\ell)(\beta_1-\beta_2).\]
	
	Note that for every $\pi_t\in\cS_n^{\ell}$, the permutation $\pi_t^c\in\cS_n^{\ell}$ and that $(\pi_t^c)^c=\pi_t$. Thus, for every $\epsilon>0$, there exists $t_{\epsilon}>t_0$ 
	such that for every $t>t_{\epsilon}$
	\begin{align*}
	\sum_{\pi_t\in\cS_n^{\ell}}\mu_{\ell}(\pi_t) C_t(\pi)&\geq \frac{1}{|\cS_n^{\ell}|} \frac{1}{2} \sum_{\pi_t\in\cS_n^{\ell}} C_t(\pi_t)+C_t(\pi_t^c)-\epsilon R
	\geq C+\frac{1}{2}\ell(a+\ell)(\beta_1-\beta_2) -\epsilon R,
	\end{align*}
	which concludes the proof.
\end{IEEEproof}

\vspace*{.1in}We can now complete the proof of Theorem \ref{thm:main1}.
\begin{IEEEproof}[Proof of Theorem \ref{thm:main1}]
	From Lemma \ref{lem:converge} and Lemma \ref{lem:uni}, we have 
	$$
C_{\rm avg}^{\wt}(\cN)\stackrel{\text{a.s.}}= \lim_{t\to\infty}\frac{1}{t}\sum_{r=1}^t\E[C_t(\cN)]\stackrel{\text{a.s.}}{=} \lim_{t\to\infty}\frac{1}{t}\sum_{r=t_0}^t\sum_{\pi\in\cS_n}\Pr(\pi_r=\pi)C_r(\pi).
	$$
	Observe that $(\cS_n^{\ell})_{\ell}$ partitions the set $\cS_n$, and we can continue as follows:
	\begin{align*}
	C_{\rm avg}^{\wt}(\cN)&\stackrel{\text{a.s.}}{=} \lim_{t\to\infty}\frac{1}{t}\sum_{r=t_0}^t\sum_{\ell=0}^{\min \{a,n_1\}}\sum_{\pi\in\cS^{\ell}_n} 
	\Pr(\pi_r=\pi |\cS_n^{\ell})\Pr(\pi_r\in\cS_n^{\ell})C_r(\pi) \\
	&= \lim_{t\to\infty}\frac{1}{t}\sum_{r=t_0}^t\sum_{\ell=0}^{\min \{a,n_1\}} \Pr(\pi_r\in\cS_n^{\ell}) \E_{\mu_{\ell}}\sparenv{C_r(\cN)}
	\end{align*}
By Lemma \ref{lem:prob1}, for every $\epsilon>0$, there is $t_{\epsilon}>t_0$ such that 
$\big|\Pr(\pi_r\in\cS_n^{\ell})-\frac{\binom{n_1}{\ell}\binom{n_2}{a-\ell}}{\binom{n}{a}}\big|\leq \epsilon$. 
Hence, for every $\epsilon>0$, 
	$$
	C_{\rm avg}^{\wt}(\cN)\stackrel{\text{a.s.}}{\geq} \lim_{t\to\infty}\frac{1}{t}
	\Big(\sum_{r=t_{\epsilon}}^{t}\sum_{\ell=0}^{\min \{a,n_1\}} \frac{\binom{n_1}{\ell}\binom{n_2}{a-\ell}}{\binom{n}{a}} \E_{\mu_{\ell}}
	[C_r(\cN)] -\sum_{r=t_0}^{t_{\epsilon}}n_1 R\Big),
	$$
	where $R=\max_{\pi_t\in\cS_n} C_t(\pi_t)$.Together with Lemma \ref{lem:perm} this yields
	\begin{equation}\label{eq:U}
C_{\rm avg}^{\wt}(\cN)\stackrel{\text{a.s.}}{\geq} C+ \sum_{\ell=0}^{n_1}\frac{\binom{n_1}{\ell}\binom{n_2}{a-\ell}}{\binom{n}{a}}\frac{\ell(a+\ell)(\beta_1-\beta_2)}{2}.
    \end{equation}	
By Lemma \ref{lem:lower_bound_iccap}, the right-hand side of this inequality gives a lower bound on capacity. It can be
transformed to the expression on the right-hand side of \eqref{eq:f} by repeated application of the Vandermonde convolution formula.
\end{IEEEproof}

Thus, we have proved that the average minimum cut (and thus, the capacity) is almost surely bounded below by an expression which is strictly 
greater than $C$, and accounting for the dynamics of the fixed-cost network enables one to support storage of a larger file than in the static
case of \cite{AkhKiaGha2010cost}.

To summarize the results of this section, we have proved that
   \begin{equation}\label{eq:max}
   \ccap(\cN)-C\stackrel{\text{a.s.}}{\geq}
   \max\Big\{\frac{n_1(n_1-1)}{2}\varepsilon_1,\frac{\beta_1-\beta_2}2\frac{an_1}{n}\Big(a+1+\frac{n_1-1}{n-1}(a-1)\Big)\Big\},
      \end{equation}
where the first of the bounds on the right is valid under assumption \eqref{as:1}.
   To give numerical examples, let us return to Example \ref{ex:basic}. 
Applying Theorem 
  \ref{thm:main1} to Example \ref{ex:basic} yields $\ccap(\cN)\geq 214\beta_2+3.7\beta_2$.
At the same time, Theorem \ref{thm:main12} states that the storage capacity is bounded below by $214\beta_2+\frac{9}{4}\beta_2$, showing that the choice of $h$ is not always optimal. Generally, the lower bound on capacity of Theorem \ref{thm:main12} is $C+\frac{n_1n_2}{2n}(\beta_1-\beta_2)$ and the bound of
Theorem \ref{thm:main1} is approximately $C+\frac{n_1a^2}{2n}$. Therefore, Theorem \ref{thm:main1} provides a better bound on the storage capacity 
when $a$ is roughly above $\sqrt{n_2}$.

Since the storage capacity can be increased while the average amount of symbols each node $v_i$ transmits is at most $\obeta_i$, after a long period 
of time (for large enough $t$), the total bandwidth that was used for repair in the dynamical model is equal to the total bandwidth that was used 
for repair in the static model. 

To conclude this section, we address the question regarding the accuracy of the derived bounds on $\E\sparenv{C_{\rm avg}^{\wt}(\cN)}.$
In the next proposition we derive an upper bound on this quantity.
	\begin{proposition}
		\label{clm:upper}
		Let $(\cN,\obeta,\obbs,\wt)$ be a storage network. We have ($\mu_1$-a.s.)
		\[C_{\rm avg}^{\wt}(\cN)\leq C+ \frac{a n_1 (a+n_1)}{2n}(\beta_1-\beta_2) .\]
	\end{proposition}

\begin{IEEEproof}
Given $\pi\in\cS_n^{\ell},$ denote by $\overline{\pi}$ the permutation in which the first $n_2-a+\ell$ positions contain nodes from $L$, the 
next $n_1-\ell$ positions contain nodes from $U$, and the last $a$ positions are the same as $\pi$.  Lemma \ref{lem:opchoose} and 
Remark~\ref{rem:opchoose} imply that $C_t(\overline{\pi}_t)\geq C_t(\pi_t)$. By \eqref{eq:equal1} and by Lemma \ref{lem:cap1} we obtain 
     $$
C_t(\overline{\pi}_t)+C_t(\overline{\pi}_t^c)= 2C+\ell(a+n_1)(\beta_1-\beta_2).
	  $$
Hence, 
	\begin{align*}
	C_t(\pi_t)+C_t(\pi_t^c) &\leq C_t(\overline{\pi}_t)+C_t(\overline{\pi}^c_t) 
	= 2C+\ell(a+n_1)(\beta_1-\beta_2)
	\end{align*}
	which implies that 
	\begin{align*}
	\sum_{\pi_t\in\cS_n}\Pr(\pi_t)C_t(\pi_t) 
	&= \sum_{\ell=0}^{\min \{a,n_1\}}\sum_{\pi_t\in\cS^{\ell}_n}\Pr(\pi_t|\cS_n^{\ell})\Pr(\cS_n^{\ell})C_t(\pi_t) \\
	&= \sum_{\ell=0}^{\min \{a,n_1\}}\Pr(\cS_n^{\ell}) \sum_{\pi_t\in\cS^{\ell}_n}\Pr(\pi_t|\cS_n^{\ell})C_t(\pi_t).
	\end{align*}
	For $t\to\infty$, $\Pr(\pi_t | \cS_n^{\ell})$ is uniform. Hence,
	\begin{align*}
	C_{\rm avg}^{\wt}(\cN)
	&= \sum_{\ell=0}^{\min \{a,n_1\}}\Pr(\cS_n^{\ell}) \Big(\frac{1}{|\cS_n^{\ell}|}\sum_{\pi_t\in\cS^{\ell}_n}C_t(\pi_t)\Big)\\
	&\leq \sum_{\ell=0}^{\min \{a,n_1\}}\Pr(\cS_n^{\ell}) \biggl(\frac{\sum_{\pi_t\in\cS^{\ell}_n}(C_t(\overline{\pi}_t)+C_t(\overline{\pi}^c_t))}{2|\cS_n^{\ell}|}\biggl)\\
	&= \sum_{\ell=0}^{\min \{a,n_1\}}\Pr(\cS_n^{\ell})\Big(C+\frac{1}{2}\ell(a+n_1)(\beta_1-\beta_2)\Big)\\
	&= C+ \sum_{\ell=0}^{n_1}\frac{\binom{n_1}{\ell}\binom{n_2}{a-\ell}}{\binom{n}{a}}\frac{\ell(a+n_1)(\beta_1-\beta_2)}{2},
	\end{align*}
	where the last equality follows from Lemma \ref{lem:prob1}.
	By Vandermonde's identity we obtain that almost surely
	\[C_{\rm avg}^{\wt}(\cN)\leq C+\frac{a n_1 (a+n_1)}{2n}(\beta_1-\beta_2).\]
\end{IEEEproof}

Proposition \ref{clm:upper} and Theorem \ref{thm:main1} jointly result in the following 
(a.s.) inequalities for 
the average cut of the fixed-cost storage network:
\begin{align} 
&\frac{an_1(\beta_1-\beta_2)}{2n}\Big(a+1+\frac{n_1-1}{n-1}(a-1)\Big) 
\leq C_{\rm avg}^{\wt}(\cN)-C
\leq \frac{an_1(\beta_1-\beta_2)}{2n}(a+n_1)\label{eq:bounds}
\end{align}
where $C$ is given in Lemma \ref{lem:mincutknown}. 
For the above example, we obtain for the gap between $ C_{\rm avg}^{\wt}(\cN)$ and $C$ an upper bound of $9.75.$ 
Generally, the difference between the upper and lower bounds (discounting the common multiplier) is $\frac{(n-a)(n_1-1)}{n-1}.$
Of course, this does not directly result in an upper bound on capacity of $\cN$, which appears to be a difficult question (a loose
upper bound was obtained in \eqref{eq:upper}, which in the example gives a gap of at most 21.5).

\section{Networks With Memory}
\label{sec:memory}
Let us assume that the data collector $DC_t$ in the dynamical fixed-cost model is aware of the state of the network;
specifically, we assume that it selects the set $D_t$ of $k'$ active nodes for data retrieval with full knowledge of the permutation $\pi_t.$
Under this assumption, $DC_t$ can choose the nodes that {\em maximize} the cut between itself and $D_t$.

With this in mind, we give the following definition. Let $(\cN,\obeta,\obbs, h)$ be a storage network and let $C_t^h(D_t)$ denote the cut at time $t$ for the selection of $D_t$ active storage nodes. Let
  \begin{equation}\label{eq:maxcut}
  C^{max,h}_t(\cN)=C^{max,h}_t(\cA_t)\triangleq \max_{D_t\subseteq \cA_t,\; |D_t|= k'} \{C_t^{h}(D_t)\}.
  \end{equation}
(cf. \eqref{eq:mincut}).
Although the memory property does not affect the storage capacity when $\beta_i=\beta_0$ for all $i\in [n]$, 
using our idea of controlling the transmission policy enables us to increase the storage capacity. As a main result of this section,
we show that the capacity of the network can be increased over the non-causal model.

Recall our notation $[n]=U\cup L,$ where $|U|=n_1,|L|=n_2.$
Throughout this section we denote $\hat{a}\triangleq k'-n_2\geq 0.$ 
The following lemma is a natural minimax analog of Lemma \ref{lem:mincutknown}. 
\begin{lemma}
	\label{lem:mincutknown2}
	Let $(\cN,\obeta,s,\wt)$ be a static fixed-cost storage network and let 
	   \begin{equation}\label{eq:maxcut1}
	C'\triangleq \min_{\pi\in\cS_n} \mathset{C^{max,h}(\pi)}=\min_{\substack{t\geq 0,\\ \os\in V^{\infty}}} \mathset{C^{max,h}_t}.
	\end{equation}
	Then
	\begin{align}
	C'&\stackrel{}{=}\sum_{i=1}^{\hat{a}}  n_1\beta_1+n_2\beta_2-i\beta_1
	\; + \sum_{j=1}^{n_2}(n_1- \hat{a})\beta_1+n_2\beta_2-j\beta_2. \label{eq:C2}
	\end{align}
\end{lemma}
Lemma \ref{lem:mincutknown2} can be obtained from the next lemma which is a modified version of Lemma \ref{lem:opchoose}, together with the fact that every permutation appears as an associated permutation in $(\cN,\obeta,\obbs)$ $\mu_1$-almost surely. 
\begin{lemma}
	\label{lem:opchoose2}
	Let $(\cN,\obeta,\os,\wt)$ be a storage network. For $t>t_0$, $C^{max,\wt}_t(\cN)$ is obtained when $D_t\supseteq L$.
\end{lemma}

The proof of Lemma \ref{lem:opchoose2} is similar to the proof of Lemma \ref{lem:opchoose} and is given in the appendix.
Note that according to Lemma \ref{lem:opchoose}, the selection that minimizes the cut at time $t$ is the node from $U$ that 
has failed before the other nodes in $U$.

\begin{remark}
	\label{rem:opchoose2}
	\normalfont
Similarly to Remark \ref{rem:opchoose}, from the proof of Lemma \ref{lem:opchoose} it follows that after choosing the nodes in $L$, we 
should choose the remaining $\hat{a}$ nodes in the order reversed from the order of their failure, starting with the most recently failed node. 
\end{remark}

For a network with memory $(\cN,\obeta,\obbs)$ we denote the average (maximum) cut and the storage capacity by $C_{\rm avg}^{max,h}, \ccap^{m}(\cN)$, respectively. 
The main result of this section is stated in the following theorem.
\begin{theorem}
	\label{thm:main2}
	Let $(\cN,\obeta,\obbs)$ be a (random) storage network with memory. We have ($\mu_1$-a.s.)
	\[\ccap^{m}(\cN)\geq C'+\frac{\beta_1-\beta_2}{2}\frac{n_1 n_2 \hat{a}}{n}\parenv{2-\frac{\hat{a}-1}{n-1}} .\]
\end{theorem}
In this section we denote by $\hat{\cS}_n^{\ell}$ the set of all permutations over $[n]$ with exactly $\ell$ elements from $U$ in the last $\hat{a}$ positions. To prove Theorem \ref{thm:main2} we need the following lemma.
\begin{lemma}
	\label{lem:perm2}
	Let $(\cN,\obeta,\obbs,\wt)$ be a storage network with memory. Let $\pi_t$ be the permutation at time $t$ and assume that $\pi_t$ is distributed uniformly over $\hat{\cS}_n^{\ell}$. We have
	    $$
	\E\sparenv{C^{max,\wt}_t(\cN)}\geq C'+\frac{1}{2}\ell \parenv{2n_2-\hat{a}+\ell}(\beta_1-\beta_2),
	   $$
where $C'$ is given in \eqref{eq:maxcut1}.	   
\end{lemma}

\begin{IEEEproof}
	For any permutation $\pi\in\hat{\cS}_n^{\ell}$, let $\overline{\pi}\in\hat{\cS}_n^{\ell}$ be a permutation in which the first $n_1-\ell$ positions contain only nodes from $U,$ and the last $\hat{a}$ positions are exactly as in $\pi$. Then by Lemma \ref{lem:opchoose2} and Remark \ref{rem:opchoose2} 
we have $C_t^m(\pi_t)\geq C_t^m(\overline{\pi}_t)$. This implies that by fixing the last $\hat{a}$ positions in $\pi_t$, we can bound 
$C_t^{max,\wt}(\pi_t)$ below by $C_t^{max,\wt}(\overline{\pi}_t)$. We claim that
	\begin{align*}
	&C_t^{max,\wt}(\pi_t)+C_t^{max,\wt}(\pi_t^c) \geq C' +\frac{1}{2}\ell(2n_2-\hat{a}+\ell)(\beta_1-\beta_2).
	\end{align*}
	Note that if $\pi_t\in\hat{\cS}_n^{\ell}$ then $\pi_t^c\in\hat{\cS}_n^{\ell}$ as well. Hence, $C_t^{max,\wt}(\pi_t)+C_t^{max,\wt}(\pi_t^c)\geq C_t^{max,\wt}(\overline{\pi}_t)+C_t^{max,\wt}(\overline{\pi}_t^c)$.  By Lemma \ref{lem:cap1} we obtain 
	\begin{align*}
	C_t^{max,\wt}(\overline{\pi}_t)
	&= C'+\sum_{v\in D_t\cap U}f_{\overline{\pi}_t}(v)(\beta_1-\beta_2) \\
	&= C'+\sum_{\substack{v\in D_t\cap U\\ (\overline{\pi}_t)^{-1}(v)\in \mathset{n-\hat{a}+1,\dots, n}}} f_{\overline{\pi}_t}(v)(\beta_1-\beta_2)
	\end{align*}
and the same holds for $\overline{\pi}_t^c$. 

Let $v\in D_t\cap U$ with $(\overline{\pi}_t)^{-1}(v)\in \mathset{n-\hat{a}+1,\dots,n}$, 
meaning that $v$ is in one of the last $\hat{a}$ positions. Let 
   $$
   b:=|L\cap \{n-\hat{a}+1,\dots, (\overline{\pi}_t)^{-1}(v)\}|.
   $$
Using definition of $\overline{\pi}_t\in\hat{\cS}_n^{\ell}$ and Lemma \ref{lem:opchoose2}, we 
now observe that $f_{\overline{\pi}_t}(v)=n_2-(\hat{a}-\ell)+b.$ 
For $\overline{\pi}_t^c$ we have 
   $$
   f_{\overline{\pi}_t^c}(v)=n_2-(\hat{a}-\ell)+(\hat{a}-\ell)-b=n_2-b.
   $$
Overall we obtain 
	\begin{align*}
	C_t^{max,\wt}(\overline{\pi}_t)+C_t^{max,\wt}(\overline{\pi}_t^c) 
	&= 2C' +\sum_{v\in D_t\cap U} \parenv{f_{\overline{\pi}_t}(v)+f_{\overline{\pi}_t^c}(v)}\parenv{\beta_1-\beta_2} \\
	&=2C'+\ell\parenv{2n_2-\hat{a}+\ell}\parenv{\beta_1-\beta_2}
	\end{align*}
	which in turn implies that 
	\begin{align*}
	C_t^{max,\wt}(\pi_t)+C_t^{max,\wt}(\pi_t^c)
	&\geq 2C'+\ell\parenv{2n_2-\hat{a}+\ell}\parenv{\beta_1-\beta_2}.
	\end{align*}
	We conclude the proof by noticing that 
	\begin{align*}
	\E\sparenv{C^{max,\wt}_t(\cN)}
	&=\sum_{\pi_t\in \hat{\cS}_n^{\ell}} \Pr(\pi_t)C_t^{max,\wt}(\pi_t) \\
	&= \frac{1}{\abs{\hat{\cS}_n^{\ell}}}\frac{1}{2}\sum_{\pi_t\in \hat{\cS}_n^{\ell}} \parenv{C_t^{max,\wt}(\pi_t)+C_t^{max,\wt}(\pi_t^c)}\\
	&\geq C'+\frac{1}{2}\ell \parenv{2n_2-\hat{a}+\ell}(\beta_1-\beta_2).
	\end{align*}
\end{IEEEproof}

We can now prove Theorem \ref{thm:main2}.
\begin{IEEEproof}[Proof of Theorem \ref{thm:main2}]
	Consider $\E[C^{max,\wt}_t(\cN)]$ and note that since $(\hat{\cS}_n^{\ell})_{\ell}$ partition the set $\cS_n$ we have
	\begin{align*}
	\E_{\mu}[C^{max,\wt}_{\rm avg}(\cN)]
	&= \sum_{\pi_t\in\cS_n}\Pr(\pi_t)C_t(\pi_t) \\
	&= \sum_{\ell=0}^{\min \mathset{\hat{a},n_1}}\sum_{\pi_t\in\cS^{\ell}_n}\Pr(\pi_t|\hat{\cS}_n^{\ell}) \Pr(\hat{\cS}_n^{\ell})C_t(\pi_t) \\
	&= \sum_{\ell=0}^{\min \mathset{\hat{a},n_1}}\Pr(\hat{\cS}_n^{\ell}) \sum_{\pi_t\in\cS^{\ell}_n}\Pr(\pi_t|\hat{\cS}_n^{\ell})C_t(\pi_t)\\
	&{\geq} \sum_{\ell=0}^{\min \mathset{\hat{a},n_1}}\Pr(\hat{\cS}_n^{\ell})\Big(C' +\frac{\ell(2n_2-\hat{a}+\ell)\beta_1}{2}
	+\frac{\ell(2n_2-\hat{a}+\ell)\beta_2}{2}\Big)\\
	&{=} C'+\sum_{\ell=0}^{\hat{a}} \frac{\binom{n_1}{\ell}\binom{n_2}{\hat{a}-\ell}}{\binom{n}{\hat{a}}} \frac{\ell(2n_2-\hat{a}+\ell)(\beta_1-\beta_2)}{2},
	\end{align*}
	where the inequality follows from Lemma \ref{lem:perm} and the last equality follows from Lemma \ref{lem:prob1} (with $\hat{a}$) and the fact that the stationary distribution of $\pi_t$ is the uniform distribution. The final expression is obtained by repeated use of the Vandermonde convolution formula. The average cut bounds the storage capacity below since we can follow the same arguments as in Lemma \ref{lem:lower_bound_iccap} with $C_t^{\max,\wt}$ instead of $C_t^{\wt}$.
\end{IEEEproof}


\vspace*{.1in}For a numerical example we return to Example \ref{ex:basic}. If at each time $t$, $DC_t$ chooses the $k'$ nodes which yield the maximum 
cut, by Theorem \ref{thm:main2}, the storage capacity is $\ccap^m(\cN)\geq C'+13\frac{1}{3}\beta_2$, where $C'=269\beta_2$. This is 
much greater than the lower bound computed earlier for the non-causal case, and in fact even breaks above the 
static-case upper bound of \eqref{eq:upper}.
 
As seen from Theorem \ref{thm:main2}, if $\beta_1=\beta_2$ the bound below is equal to the storage capacity of the static model. 
This comes as no surprise since the network is invariant under permutations of the storage nodes.

\section{Extensions: Different failure probabilities}
\label{Sec:ext}
The dynamical model that we studied so far assumes that all the storage nodes have the same probability of failure. In reality, this may not 
be the case. An interesting extension of the above results would address a fixed-cost model in which the failure probability depends on the 
node (or a group to which the node belongs). We immediately note that the assumption of different failure probabilities does not affect the storage 
capacity of the static fixed-cost model.

Switching to the dynamical models, let us first assume that nodes from $L$ fail independently with probability $p$ and nodes from $U$
fail (independently) with probability $q$. It is possible to adjust the weight function used in Section \ref{sec:static} to prove a lower
bound on the network capacity.
As before, let 
$(\cN,\obeta,\obbs)$ be a fixed-cost storage network with $n$ storage nodes and let $|U|=n_1,|L|=n_2$. 
For $\os_t\in U$ we put $h_t(v_i)=h_U(v_i)$ and for $\os_t\in L$ we put $h_t(v_i)=h_L(v_i),$ where
\begin{align*}
h_U(v_i)&=\begin{cases}
\beta_1+\varepsilon_1 & v_i\in U\setminus \os_j \\
\beta_2 & v_i\in L \\
0 & v_i=\os_j,
\end{cases}
\end{align*}
and 
\begin{align*}
h_L(v_i)&=\begin{cases}
\beta_1-\frac{q(n_1-1)}{pn_2}\varepsilon_1 & v_i\in U \\
\beta_2 & v_i\in L\setminus \os_j \\
0 & v_i=\os_j
\end{cases}
\end{align*}
and $0\le \varepsilon_1\le \beta_1.$
By a calculation similar to Lemma \ref{lem:constsat} it is straightforward to check that the constraints given by $\obeta$ are satisfied. 
Moreover, the proof of Theorem \ref{thm:main12} does not use the fact that the stationary distribution of the associated permutations is 
uniform. Thus, from Lemma \ref{lem:opchoose} and Lemma \ref{lem:cap1} we obtain the following statement.
\begin{theorem}
	\label{cor:1}
Let $(\cN,\obeta,\obbs,h)$ be a fixed-cost storage network with weight function $h$ as defined above. Assume that the failure probability of a node 
from $U$ is $q>0$ and of a node from $L$ is $p>0$. Fix $\varepsilon_1>0$ such that $\beta_1-\beta_2\geq \frac{qn(n_1-1)}{pn_2}\varepsilon_1$. The 
storage capacity is bounded below by
	\begin{equation}\label{eq:pq}
\ccap(\cN)\stackrel{a.s.}{\geq} C+\frac{n_1(n_1-1)}{2}\varepsilon_1,
    \end{equation}
	where $C$ is given in Lemma \ref{lem:mincutknown}.
\end{theorem}
For a numerical example consider Example \ref{ex:basic} with $q=\frac1{40}$ and $p=\frac{3}{40}$. 
Let us choose $\varepsilon_1=\frac{pn_2}{qn(n_1-1)}\beta_2=\frac{1}{6}\beta_2$. From \eqref{eq:pq} we now obtain
	$$
	\ccap(\cN)\geq (214+7.5)\beta_2,
	$$
where as above, $C=214\beta_2$ is the value of the min-cut in the static case.
As above in this paper, the assumption on $\varepsilon_1$ introduced in the theorem limits the increase of the network capacity. Lifting the assumption
suggests following the path taken in Theorem~\ref{thm:main1} of Sec.~\ref{sec:general}. To implement this idea, we need to find the
stationary distribution of the Markov random walk on $\cS_n$ that arises under our assumption. This is however not an easy task, and the classic (asymptotic) results such as in \cite{FOP85} seem not to be of help here.
We have succeeded to perform the analysis in the simple case of $n=n_2+1$, i.e., of the ``upper'' set formed of a single node $U=\{u\}$, and
we present this result in the remainder of this section.

Suppose that the failed nodes in the sequence $\bbs$ are chosen independently and that $\Pr(\bbs_i=v)=p$ if $v\in L$ and $\Pr(\bbs_i=v)=q$ 
if $v\in U.$ Assuming that $p,q\neq 0$, almost surely there exists a finite time $t_0$ such that all the nodes have failed at least once by 
$t_0.$ 
Choosing the next failed node gives rise to a permutation on $\cS_n,$ and the conditional probabilities $\Pr(\pi_t | \pi_{t-1})$
between the permutations are well defined and can be found explicitly. 
The probabilities  $\Pr(\pi_t | \pi_{t-1})$ define an ergodic Markov chain with a 
unique stationary distribution $\nu$. 

Define a partition of $\cS_n$ into $n$ blocks $P_i,i\in[n]$. Let $\pi\in P_i$ if and only if $\pi^{-1}(u)=i$ \bcomment{where $u$ is the (unique) node in $U$}. The partition $(P_i)$ 
defines an obvious equivalence relation on $\cS_n$, and $|P_i|=(n-1)!$ for all $i$.

It turns out that the stationary probabilities of equivalent permutations are the same, i.e., $\nu(\pi)$ depends only on the block $P_i\ni \pi.$ The distribution $\nu$ is given in the next lemma.

For any real number $r$ and natural number $k$ we define $\binom{r}{k}=\frac{r(r-1)\dots (r-k+1)}{k!},$ and put $\binom{r}{0}=1$.
\begin{lemma}
	\label{lem:difprob1}
	Let $(\cN,\obeta,\obbs)$ be a dynamical storage network with $n=n_1+n_2$ nodes, where $n_1=1$. 
Let $0<q\leq p$ and suppose that $\bbs_i, i=1,2,\dots$ 
are independent random variables with $\Pr(\bbs_i=v)=p$ if $v\in L$ and $\Pr(\bbs_i=v)=q$ if $v\in U$. 
Let $\pi\in P_i$ and define the distribution 
	$$
	\nu(\pi)=\frac{1-q}{(n-1)!}\binom{\frac{1}{p}-1}{n-2}^{-1}\binom{\frac{1}{p}-n-1+i}{i-1}.
	$$
Then $\nu$ is the stationary distribution of the Markov chain with state space $\cS_n.$
\end{lemma}

\begin{IEEEproof}
1.	We first note that for any $t$, $\Pr\parenv{\pi_{t+1} |\pi_{t}}=q$ if $\pi_{t+1}\in P_n$ and $\Pr\parenv{\pi_{t+1} |\pi_{t}}=p$ otherwise. This implies that for a fixed $\pi_t$, 
	\[(n-1)p+q=\sum_{i=1}^n \Pr\parenv{\pi_{t+1}\in P_i | \pi_t}=1.\]
	Hence, $\frac{1}{p}\geq n-1$ which implies that all the binomial coefficients in $\nu(\pi)$ are positive. Moreover, since $(n-1)p=1-q$ we obtain that if $\pi\in P_n$ then the expression for $\nu(\pi)$ simplifies as follows
	\begin{align}
	\nu(\pi)&=\frac{1-q}{(n-1)!}\binom{\frac{1}{p}-1}{n-2}^{-1}\binom{\frac{1}{p}-n-1+n}{n-1}\notag\\
	&=\frac{1-q}{(n-1)!}\frac{\frac{1}{p}-n+1}{n-1}\notag\\
	&=\frac{q}{(n-1)!}. \label{eq:nu-n}
	\end{align}

2. Let us check that $\nu$ is a probability vector. 
As already remarked, $\nu(\pi)>0$ for all $\pi\in \cS_n$. Obviously,
if $\pi,\sigma\in P_i$ for some $i$, then $\nu(\pi)=\nu(\sigma)=\frac{1}{(n-1)!}\nu\parenv{\mathset{P_i}}$. 
	
By the definition of $\nu$ we have that $\nu(\{P_{i+1}\})=\nu(\{P_i\})(1+\frac{1-pn}{pi})$ for all $i\le n-1.$ Therefore,
	\begin{align*}
	\sum_{\pi\in\cS_n}\nu(\pi)
	&=\sum_{i=1}^n \nu\parenv{\mathset{P_i}} \\
	&= \sum_{i=1}^{n-1} \nu\parenv{\mathset{P_i}} + \nu(\mathset{P_n}) \\
	&= \nu(\{P_1\})\Big(1+\sum_{j=1}^{n-2}\prod_{i=1}^j \Big(1+\frac{1-pn}{pi}\Big)\Big) +q.
	\end{align*}
	Note that 
	\[\prod_{i=1}^j \parenv{1+\frac{1-pn}{pi}}=\binom{j+\frac{1-pn}{p}}{j}\]
which implies that 
	$$
\sum_{j=1}^{n-2}\prod_{i=1}^j \Big(1+\frac{1-pn}{pi}\Big)=\binom{\frac{1}{p}-1}{n-2}-1.
   $$
Since for $\pi\in P_1$, $\nu\parenv{\mathset{P_1}}=(n-1)!\nu(\pi)$ and $\nu(\pi)=\frac{1-q}{(n-1)!}\binom{\frac{1}{p}-1}{n-2}^{-1}$, we have 
	\[\sum_{\pi\in\cS_n}\nu(\pi)=(n-1)!\binom{\frac{1}{p}-1}{n-2}\nu(\pi)+q=1.\]
	
Finally let us show that $\nu$ is a stationary vector of the transition matrix. Fix $t$ and consider the sum $\sum_{\pi\in\cS_n}\nu(\pi)\Pr\parenv{\pi_{t+1}=\sigma | \pi_t=\pi}$. For $\sigma\in P_i$, this sum has exactly $n$ non-zero terms, 
of which 
$i$ are for $\pi\in P_{i+1}$ and $n-i$ for $\pi\in P_{i}$. 
Therefore,  
if $\sigma\in P_i$, we obtain 
	\begin{align*}
&\sum_{\pi\in\cS_n}\nu(\pi)\Pr\parenv{\pi_{t+1}=\sigma | \pi_t=\pi}
=\frac{p i}{(n-1)!} \nu(\mathset{P_{i+1}})+\frac{p(n-i)}{(n-1)!}\nu(\mathset{P_i}).
	\end{align*}
Since $\nu\parenv{\mathset{P_{i+1}}}=\nu\parenv{\mathset{P_i}}(1+\frac{1-pn}{pi})$ for $i\le n-1$, we have  
	\begin{align*}
\sum_{\pi\in\cS_n}\nu(\pi)\Pr\parenv{\pi_{t+1}=\sigma | \pi_t=\pi}
    &= \frac{p}{(n-1)!}\nu(\mathset{P_i})\parenv{i\parenv{1+\frac{1-pn}{pi}}+(n-i)}\\
	&=\frac{1}{(n-1)!}\nu(\mathset{P_i})\\
	&=\nu(\sigma).
	\end{align*}
Now assume that $\sigma\in P_n$. We obtain 
	\begin{align}
	\label{eq:11}
	\sum_{\pi\in\cS_n}\nu(\pi)\Pr(\pi_{t+1}=\sigma | \pi_t=\pi)
	&= \frac{1}{(n-1)!}\sum_{i=1}^n q\nu(\mathset{P_i}) \\ \nonumber
	&= \frac{q}{(n-1)!}\parenv{\sum_{i=1}^{n-1} \nu(\mathset{P_i}) + \nu(\mathset{P_n})}.
	\end{align}
	Using the fact that $\sum_{i=1}^{n-1} \nu(\mathset{P_i})=1-q$ jointly with \eqref{eq:11}, we conclude that  
	\begin{align*}
	\sum_{\pi\in\cS_n}\nu(\pi)\Pr(\pi_{t+1}=\sigma| \pi_t=\pi)
	&= \frac{q}{(n-1)!}\parenv{1-q+q},
	\end{align*}
recovering \eqref{eq:nu-n}. This concludes the proof.
\end{IEEEproof}

We can now bound the storage capacity below. Recall that $\cS_n^{\ell}$ denotes the set of all permutations on $[n]$ with $\ell$ numbers from $
[n_1]$ in the last $a$ positions. 
Let us use Lemmas \ref{lem:cap1} and \ref{lem:difprob1} to calculate the expected minimum cut. In the below calculation we are taking
some liberty in dealing with the conditional distribution $\nu(\pi_t|P_i)$ which operationally is the limiting (conditional) probability on $
\cS_n.$ A more rigorous approach requires defining a conditional distribution for a finite time $t$ and arguing that it approaches $\nu(\pi_t|P_i)=
\frac1{(n-1)!}.$ At the same time, the final answer below is correct as written. 
We proceed as follows:
	\begin{align*}
	\sum_{\pi_t\in\cS_n}\nu(\pi_t)C_t^{\wt}(\pi_t)&= \sum_{i\in [n]}\sum_{\pi_t\in P_i}\nu(P_i)\nu(\pi_t|P_i)C_t^{\wt}(\pi_t)\\
	&= \sum_{i\in [n-a]}\sum_{\pi_t\in P_i}\nu(P_i)\nu(\pi_t|P_i)C_t^{\wt}(id)\\
	&\quad+
	\sum_{i=n-a+1}^n \frac{(1-q)}{(n-1)!}\binom{\frac{1}{p}-1}{n-2}^{-1}\binom{\frac{1}{p}-n+i-1}{i-1}\sum_{\pi_t\in P_i}C_t^{\wt}(\pi_t)\\
	&= (1-q)\binom{\frac{1}{p}-1}{n-2}^{-1}\binom{\frac{1}{p}-a}{n-a} C \\
	&\quad + \sum_{i=n-a+1}^n (1-q)\binom{\frac{1}{p}-1}{n-2}^{-1}\binom{\frac{1}{p}-n+i-1}{i-1} (C+(i-n+a)(\beta_1-\beta_2))
	\end{align*}
	where $C$ is given \eqref{eq:C}.
To argue that this expression can be used in the lower bound on $\ccap{(\cN)}$ similar to the bound in Theorem \ref{thm:main1} (or in \eqref{eq:U})
we can repeat the arguments used in the proof of Lemma \ref{lem:perm}. Then a modified version of \eqref{eq:U} together with the above
expression for $\nu$ gives a lower bound on the capacity.

To give a numerical example, assume that we have $n=20$ with $n_2=19$, $p=\frac{4}{95}$ and $q=\frac{1}{5}$. Assume also that $\beta_1=2\beta_2$ and $k'=13$ (which implies that $a=12$). According to Lemma \ref{lem:mincutknown}, the capacity in the static model is $C=150\beta_2$. Using
the results in this section, we obtain that a.s.
	$$
\ccap(\cN)\geq (0.022\cdot150+155.4)\beta_2=158.7\beta_2.
	$$
Lemma \ref{lem:upper_bound_iccap} implies that $\ccap(\cN)\le177.45\beta_2$ and thus in the above example
we have obtained the capacity increase of more that $30\%$ of the gap between the bounds.

\section{Concluding Remarks}
\label{sec:conc}
In this work we introduced a dynamical model for distributed storage systems. 
We provided lower bounds on the capacity for the fixed-cost model with no memory and with memory. For the memoryless network, we also provided a 
simple transmission protocol that increases the storage capacity of the network over the static case. We did not manage to optimize
the weight assignment, which is left as an open question for future work, or to provide explicit code families that support reliable file storage while accounting for the time evolution of the storage network.
Another extension that we did not address is to consider more than two clusters of nodes with different values of (average) repair bandwidth.
It is also possible to argue that once the node has been repaired, it is less likely to fail for a certain period of time. At this point we do not have an approach to the analysis of this general question.

\appendix
\subsection{Proof of Lemma
\ref{lem:equalavg}}\label{app:equalavg}
	First note that Lemmas \ref{lem:converge} and \ref{lem:uni} imply that if $(\cN_2,\obeta,\obbs,\wt)$ is a random discrete-time storage network, 
	then $C_{\rm avg}^{\wt}(\cN_2)$ is almost surely a constant, which is equal to $\frac{1}{n!}\sum_{\pi\in\cS_n}C_t(\pi)$. On the other hand, if 
	$(\cN_1,\obeta,\obbs,\ot,\wt)$ is a continuous-time storage network, then we can write
	\begin{align}
	\label{eq:endi1}
	C_{\rm avg}^{\wt}(\cN_1)&=\limsup_{\tau\to\infty} \frac{1}{\tau}\int_{0}^{\tau}C_t^{\wt}(\cN_1)\mathrm{d}t \\ \nonumber
	&= \limsup_{\tau\to\infty} \frac{1}{\tau}\int_{t_0}^{\tau}\sum_{\pi\in\cS_n}\mathbbm{1}_{\pi}(\pi_t)C_t^{\wt}(\cN_1)\mathrm{d}t \\ \nonumber
	&= \limsup_{\tau\to\infty} \frac{1}{\tau}\sum_{\pi\in\cS_n} \int_{t_0}^{\tau}\mathbbm{1}_{\pi}(\pi_t)C_t^{\wt}(\pi)\mathrm{d}t
	\end{align} 
	where $t_0$ is the first time instance by which all the nodes have failed. Moreover, since $C_t^{\wt}(\pi)$ is a function of $\pi$ and not a function of $t$ we denote $C_t^{\wt}(\pi)$ by $C^{\wt}(\pi)$, and obtain
	\begin{align*}
	C_{\rm avg}^{\wt}(\cN_1)&= \sum_{\pi\in\cS_n} \limsup_{\tau\to\infty} \frac{1}{\tau}\int_{t_0}^{\tau}\mathbbm{1}_{\pi}(\pi_t)C^{\wt}(\pi)\mathrm{d}t \\
	&= \sum_{\pi\in\cS_n} \lim_{\tau\to\infty}\frac{1}{\tau}C^{\wt}(\pi)\int_{t_0}^{\tau}\mathbbm{1}_{\pi} (X(t')) \mathrm{d}t \\
	&\stackrel{a.s}{=} \sum_{\pi\in\cS_n}\frac{C^{\wt}(\pi)}{n\lambda \E\sparenv{(\pi\to\pi)}}
	\end{align*}
	where the last equality follows from \eqref{eq:22}.
	Since $\E\sparenv{(\pi\to\pi)}$ does not depent on $\pi\in\cS_n$ we obtain that $\frac{1}{n\lambda\E\sparenv{(\pi\to\pi)}}=\frac{1}{n!}$, 
	which in turn implies that $C_{\rm avg}^{\wt}(\cN_1)=C_{\rm avg}^{\wt}(\cN_2)$ almost surely. 

\subsection{Proof of Lemma \ref{lem:opchoose2}}
	Assume that $\pi_t$ is a fixed permutation and consider the information flow graph $\cX_t$. We consider a $k'$-step procedure which in each step  selects one node from $\cA_t$. Let $t'\leq t$ and assume the node $v_{i_t}^t\in \cA_t$ was selected. The cost it entails is defined as the added weight values of the in-edges of $CU_t$ that are not out-edges of previously selected nodes. Our goal is to select $k'$ nodes that maximize the cut for $\pi_t$.
	
	In order to simplify notation, we write $\pi_t=(u_1,u_2,\dots,u_n)$, i.e., $u_l=v_{\pi_t(l)}$ is the storage node that appears in the $l$th position in $\pi_t$. Moreover, with a slight abuse of notation, if $u_j$ failed at time $t'$ we will write $h_{j}(u_i)$ instead of $h_{t'}(u_i)$. 
	For $\kappa\leq k'$, consider the sub-problem at step $\kappa-1$, where the $DC_t$ has already chosen $\kappa-1$ nodes $(u_{i_1},\dots, u_{i_{\kappa-1}})$ and we are to choose the last node. Assume that the chosen nodes are ordered according to their appearance in the permutation, i.e.,  $i_1\leq i_2\leq\dots\leq i_{\kappa-1}$. Let  $u_{j_1},\dots,u_{j_m}\in L$ be nodes that were not selected up to step $\kappa-1$, i.e., 
	\[\{u_{j_1},\dots, u_{j_{m}}\}\cap \{u_{i_1},\dots, u_{i_{\kappa-1}}\}=\emptyset,\] 
	and assume also that $j_1\leq j_2\leq \dots \leq j_m$. We show that choosing $u_{j_1}$ accounts for the maximum cut. 
	
	First, we show that choosing $u_{j_1}$ maximizes the cut over all other nodes from $L$. 
Denote by $C_{\kappa-1}$ the total cost (or the cut) in step $\kappa-1.$ 
Fix $2\leq \ell\in[m]$ and note that since $j_1\leq j_{\ell}$ we may write 
   $$
i_1\leq \dots \leq i_{r_1}\leq j_1\leq i_{r_1+1}\leq \dots\leq i_{r_{\ell}}\leq j_{\ell}\leq i_{\ell+1}\leq \dots,
   $$
where $j_1$ could also be $1$. Let $C(j_1)$ be the cut value if $DC_t$ chooses $u_{j_1}$ in the $\kappa$th step, respectively. 
The change from $C_{\kappa-1}$ is formed of the following
components. First, we add the values of all the edges from $U$ to $u_{j_1}$ and from $L\{u_{j_1}\}$ to $u_{j_1}$, accounting
for $(n_1-1)(\beta_1+\varepsilon_1)+n_2\beta_2$ symbols. 
	Next, for each node $u_{i_q}$ with $r_1<q\leq \kappa-1$, we subtract $\wt_{i_q}(u_{j_1})$ from the cut value.
Overall we obtain 
	\begin{align}
		\label{eq:cj11}
		C(j_1)&=C_{\kappa-1}+n_1\beta_1+(n_2-1)\beta_2-\sum_{q=1}^{r_1}\wt_{j_1}(u_{i_q})
		-\sum_{q=r_1+1}^{\kappa-1}\wt_{i_q}(u_{j_1}).
	\end{align}
	Following the same steps for $C(j_{\ell}),\ell\ge 2$ we obtain 
	\begin{align*}
	C(j_{\ell})&=C_{\kappa-1}+n_1\beta_1+(n_2-1)\beta_2-\sum_{q=1}^{r_{\ell}}\wt_{j_{\ell}}(u_{i_q})
	- \sum_{q=r_{\ell}+1}^{\kappa-1}\wt_{i_q}(u_{j_{\ell}}).
	\end{align*}
	Since $u_{j_1},u_{j_{\ell}}\in L$, we obtain that $\wt_{j_{\ell}}(u_{i})=\wt_{j_1}(u_{i})$ and $h_{i}(u_{j_1})=h_{i}(u_{j_{\ell}})$ for all $i\in [n]$, we have
	  $$
C(j_1)-C(j_{\ell})= \sum_{q=r_1+1}^{r_{\ell}} (\wt_{j_{\ell}}(u_{i_q}) -\wt_{i_q}(u_{j_1})).
	  $$
For $u_{i_q}\in U$, we obtain $h_{j_{\ell}}(u_{i_q}) -h_{i_q}(u_{j_1})=\beta_1-\beta_2\geq 0$ and for 
$u_{i_q}\in L$, we obtain 
	$h_{j_{\ell}}(u_{i_q}) -h_{i_q}(u_{j_1})=\beta_2-\beta_2=0$, and so
   $$
C(j_1)-C(j_{\ell}) \geq 0.
   $$

Now we show that $u_{j_1}$ maximizes the cut over the selection of any node $u_{j_{\ell}}$ from $U$. We divide the argument into $2$ cases:
\begin{enumerate}
	\item Assume that $j_{\ell}< j_1$. Denote by $(i_1,\dots,i_{r_{\ell}},j_{\ell},i_{r_{\ell}+1},\dots,i_{r_1},j_1,\dots)$ the selected nodes and let $C(j_{\ell}),C(j_1)$ be the cut values if we choose $u_{j_{\ell}},u_{j_1}$, respectively. We have
	\begin{align*}
	C(j_{\ell})&=C_{\kappa-1}+(n_1-1)\beta_1+n_2\beta_2 
	- \sum_{q=1}^{r_{\ell}}\wt_{j_{\ell}}(u_{i_q})-\sum_{q=r_{\ell}+1}^{\kappa-1}\wt_{i_q}(u_{j_{\ell}})
	\end{align*}
	Subtracting $C(j_{\ell})$ from $C(j_1)$ and using \eqref{eq:cj11}, we obtain
	\begin{align*}
	\nonumber
	C(j_1)-C(j_{\ell})	&= \beta_1-\beta_2  +\sum_{q=1}^{r_{\ell}} \wt_{j_{\ell}}(u_{i_q}) -\sum_{q=1}^{r_{1}} \wt_{j_1}(u_{i_q}) 
	+ \sum_{q=r_{\ell}+1}^{\kappa-1} \wt_{i_q}(u_{j_{\ell}}) - \sum_{q=r_{1}+1}^{\kappa-1} \wt_{i_q}(u_{j_{1}})\\
	&\ge \sum_{q=1}^{r_{\ell}} \parenv{h_{j_{\ell}}(u_{i_q})-h_{j_1}(u_{i_q})} -\sum_{q=r_{\ell}+1}^{r_{1}} h_{j_1}(u_{i_q})
	\\&\hspace*{.2in}+\sum_{q=r_{\ell}+1}^{r_1} h_{i_q}(u_{j_{\ell}})+\sum_{q=r_{1}+1}^{\kappa-1} \parenv{h_{i_q}(u_{j_{\ell}})- h_{i_q}(u_{j_{1}})}\\
	&\ge \beta_1-\beta_2+\sum_{q=r_{\ell}+1}^{r_1} (h_{i_q}(u_{j_{\ell}})- h_{j_1}(u_{i_q}))
	\end{align*} 
%
%
Since $u_{i_q}\in U$ we have $h_{i_q}(u_{j_{\ell}})-h_{j_1}(u_{i_q})=0$ and for $u_{i_q}\in L$ we have 
	$h_{i_q}(u_{j_{\ell}})-h_{j_1}(u_{i_q})>0,$ we conclude that
$C(j_1)-C(j_{\ell})\geq 0$.

	\item Now assume $j_{\ell}> j_1$. This case is symmetric to the case $j_{\ell}<j_1$ and relies on the same analysis. 
We omit the details.
\end{enumerate}

According to the principle of optimality \cite[Ch. 1.3]{Ber2005}, every optimal policy consists only of optimal sub-policies,
and therefore we first need to choose all the nodes from $U$ and then choose nodes from $L$. This completes the proof.

\bibliographystyle{IEEEtranS}
\bibliography{AddlRefs,allbib}
\end{document}